\documentclass[twocolumn,twocolappendix]{aastex63}
\shorttitle{short GRB jet delay times}
\shortauthors{Beniamini et al.}

\newcommand{\swift}{\emph{Swift}}

\begin{document}
	\title{Ready, set, launch: time interval between BNS merger and short GRB jet formation}
	\correspondingauthor{Paz Beniamini}
	\email{paz.beniamini@gmail.com}
	\author{Paz Beniamini}
	\affiliation{Division of Physics, Mathematics and Astronomy, California Institute of Technology, Pasadena, CA 91125, USA}
	\author{Rodolfo Barniol Duran}
	\affiliation{Department of Physics and Astronomy, California State University, Sacramento, 6000 J Street, Sacramento, CA 95819, USA}
	\author{Maria Petropoulou}
	\affiliation{Department of Astrophysical Sciences, Princeton University, 4 Ivy Lane, Princeton, NJ 08544, USA}
	\author{Dimitrios Giannios}
	\affiliation{Department of Physics and Astronomy, Purdue University, 525 Northwestern Avenue, West Lafayette, IN 47907, USA}
	
	\begin{abstract}
		The joint detection of GW~170817/GRB 170817 confirmed the long-standing theory that binary neutron star mergers produce short gamma-ray burst (sGRB) jets that can successfully break out of the surrounding ejecta. At the same time, the association with a kilonova provided unprecedented information regarding the physical properties (such as masses and velocities) of the different ejecta constituents. Combining this knowledge with the observed luminosities and durations of cosmological sGRBs detected by the Burst Alert Telescope (BAT) onboard the \emph{Neil Gehrels Swift Observatory}, we revisit the breakout conditions of sGRB jets. Assuming self-collimation of sGRB jets does not play a critical role, we find that the time interval between the binary merger and the launching of a typical sGRB jet is $\lesssim0.1$~s. We also show that for a fraction of at least $\sim 30\%$ of sGRBs, the usually adopted assumption of static ejecta is inconsistent with observations, even if the polar ejecta mass is an order of magnitude smaller than the one in GRB 170817. Our results disfavour magnetar central engines for powering cosmological sGRBs, limit the amount of energy deposited in the cocoon prior to breakout, and suggest that the observed delay of $\sim 1.$7~s in GW 170817 /GRB 170817 between the gravitational wave and $\gamma$-ray signals is likely dominated by the propagation time of the jet to the $\gamma$-ray production site.
	\end{abstract}
	
	
	\section{Introduction}
	Multi-messenger astronomy has experienced a profound step forward with the observations of the binary neutron-star (BNS) merger event, GW~170817, in both gravitational and electromagnetic waves \citep{GW170817}. The detection of a short Gamma-ray Burst (sGRB; \citealt{Nakar2007,Berger2014}), GRB 170817, from a BNS merger has renewed the community's interest in these enigmatic explosions (see e.g. \citealp{Nakar2019} for a recent review on sGRBs from BNS mergers). GRB 170817 has forced us to revisit several important properties of GRB jets, such as their angular structure (e.g. \citealt{lamb2017,Granot2017ApJ...850L..24G,Kathirgamaraju2018,Beniamini2020}), as well as possible implications for some as-of-yet mysterious properties of `standard' GRB afterglow observations such as X-ray plateaus \citep{Oganesyan2019,BDDM2019}. More importantly, it has highlighted our need to understand how jets propagate through external media.
	
	As jets propagate out of the central engine of the sGRB, they interact with ejecta made of material launched dynamically during the compact binary merger as well as ejecta driven by the neutrinos released from the neutron star or the accretion disk formed post merger. 
	The sGRB jet propagation and ejecta interaction (possibly also determining their angular structure) has been studied
	numerically in numerous works \citep{Aloy2005,Nagakura2014,Just2016,Lazzati2017,Xie2018,Geng2019,Gill+19,Salafia2019,Kathirgamaraju2019}. Such studies are inherently complex, as the relativistic nature of the outflow naturally leads to a large range of temporal and spatial scales. Analytically, the situation may be significantly simplified by considering limiting cases for the dynamics of the ejecta, being either static \citep{Begelman1989,Marti1994,Matzner2003,Bromberg2011} or homologously expanding \citep{Duffell2018}.
	
	Comparison of model predictions with observed data can help determine the physical properties of break out, such as the time it takes the jet to break through the ejecta and the time interval between the BNS merger and the launching of the GRB jet. Similar techniques have been employed successfully in the past, mainly for long GRBs breaking out of their surrounding stellar envelopes, for which the static ejecta limit naturally applies \citep{Bromberg2012,Sobacchi2017,Petropoulou2017}.
	Previous studies comparing sGRB data to theory have focused mainly on the static limit by employing either a limited data-set of sGRBs with both measured luminosities and durations \citep{MB2014,MB2017} or a significantly more expanded data-set, but with durations only \citep{Moharana2017}.
	
	Pinning down the time-scales involved in formation and breakout of the jet is at the intersection of several key fields of current study, such as: jet formation and propagation, the nature of the central engine (and possibly constraints on the neutron star equation of state, e.g. \citealt{Lazzati2019}) and the properties of the radioactive ejecta that may be the dominant source of $r$-process production in the Universe (see \citealt{Hotokezaka2018} for a recent review). 
	
	We show here that the static versus homologous expansion limits for the ejecta propagation can be smoothly combined to form a description that holds also for intermediate situations in terms of the time delay between ejecta and jet launching and intermediate velocities of the ejecta \citep[see also][]{Hamidani2020,Lyutikov2020}.
	We then use the current sample of sGRBs with redshift determination (for which the luminosity and duration can be well determined) to place statistical constraints on the time interval between the moment of the BNS merger and the launching of the GRB jet and on the time it takes the jet to break out of the ejecta.  
	
	The paper is organized as follows. In \S \ref{sec:sample} we introduce the sample of sGRBs considered in this work. In \S \ref{sec:jbtimes} we introduce the two limiting cases (\S \ref{sec:static},\ref{sec:homologous}) for calculating the properties of jet breakout that have previously been considered in the literature. We then present a treatment that smoothly connects the two regimes (\S \ref{sec:general}) and show how sGRBs with known durations and luminosities can be used to infer physical properties of the ejecta with respect to the jets. We discuss a variety of implications of these results in \S \ref{sec:discuss} and finally conclude in \S \ref{sec:conclude}.
	
	Throughout the paper, we adopted a cosmology with $\Omega_{\rm M}=0.31$, $\Omega_\Lambda=0.69$,  and $H_0=69.6$~km s$^{-1}$ Mpc$^{-1}$.
	
	\section{Sample}
	\label{sec:sample}
	We use publicly available data from the GRB  archive\footnote{\url{https://swift.gsfc.nasa.gov/archive/grb_table/}} of the \emph{Neil Gehrels Swift Observatory} \citep{2004ApJ...611.1005G}. We select sGRBs (i.e., burst with observed $T_{90}<2$~s) detected by the \swift \, Burst Alert Telescope (BAT) from 2005 to 2019 with redshift information (either spectroscopic or photometric). Our sample consists of 27 bursts ($\sim1/4$ of the \swift-BAT sGRB sample). To estimate the isotropic $\gamma$-ray luminosity we use the BAT fluence, $\Phi$, in the 15--150 keV energy range
	\begin{eqnarray}
	L_{\gamma, \rm iso} =  \frac{4\pi d^2_L(z) \Phi}{T_{90}} \frac{\int_{1 \ {\rm keV}}^{10 \ {\rm MeV}}{\rm d} E \, E N(E)}{\int_{(1+z)15 \ {\rm keV}}^{(1+z)150 \ {\rm keV}}{\rm d}E \,E N(E)},
	\end{eqnarray}
	where $d_L(z)$ is the luminosity distance of a burst at redshift $z$ and $N(E)$ is the differential photon spectrum considered in the 1 keV--10 MeV energy range, and described by the so-called Band function \citep{Band1993} with $\alpha=-0.5$, $\beta=-2.25$, and rest-frame peak energy $E_{\rm p}=800$~keV \citep{Nava2011}.
	
	We also compare the results we derive from our \swift \, sGRB sample to the first GRB to be detected in GWs, namely GRB 170817. Since this burst was preceded by a GW trigger, it enabled the detection of a very weak prompt GRB signal with no afterglow signal until days after the event; if there was no GW trigger (i.e., under regular circumstances) there could not have been a redshift determination. Furthermore, for the purposes of this study, we are interested in the luminosities of GRBs along their jet cores. Since GRB 170817 was detected off-axis, its core luminosity is very poorly constrained \citep{Troja2019}. For these reasons, we do not include GRB~170817 in our analysis of deriving upper limits on the time interval between the moment of the BNS merger and the launching of the GRB jet, but return to discuss some specific implications for GRB 170817 in \S \ref{sec:discuss}. We use the duration data from \cite{Goldstein2017} and the constraints on the on-axis luminosity from \cite{Troja2019}, accounting for a typical on-axis efficiency, $\eta_{\gamma}\approx 0.15$ seen in other cosmological sGRBs (see definition in \S \ref{sec:static}), when discussing this specific burst.
	\section{Jet breakout times}
	\label{sec:jbtimes}
	The breakout of the jet through the BNS merger ejecta involves three dynamical timescales, namely the time interval between the BNS merger and the launch of the jet, also referred to here as the `waiting time' ($t_{\rm w}$), the  duration of the jet engine operation ($t_{\rm e}$), and the time it takes a GRB jet to break out from the BNS merger ejecta ($t_{\rm j,b}$).
	Since the ejecta is launched dynamically during the BNS merger, it is launched within several milliseconds from the moment of the merger. As this timescale is much shorter than any of the other timescales of interest considered in this situation, the launching of the ejecta can be considered as concurrent with the BNS merger.
	The breakout time $t_{\rm j,b}$ has been calculated in the limit of:
	\begin{itemize}
		\item {\it Static ejecta.} In this limit, which applies when $t_{\rm j,b},t_{\rm e}\lesssim t_{\rm w}$, the merger ejecta can be considered to be roughly static throughout the break out (see e.g. \citealt{Begelman1989,Marti1994}). 
		\item {\it Homologous ejecta expansion}. In this limit, which is relevant when $t_{\rm j,b},t_{\rm e}\gtrsim t_{\rm w}$, the evolution becomes self-similar, namely the jet breakout time is proportional to the engine timescale up to some dimensionless number that is a function of the jet's total energy. In this situation, jets typically break out more easily from the ejecta  \citep[][henceforth denoted as D18]{Duffell2018}.
	\end{itemize}
	
\begin{figure}
		\center
		\includegraphics[width=0.45\textwidth]{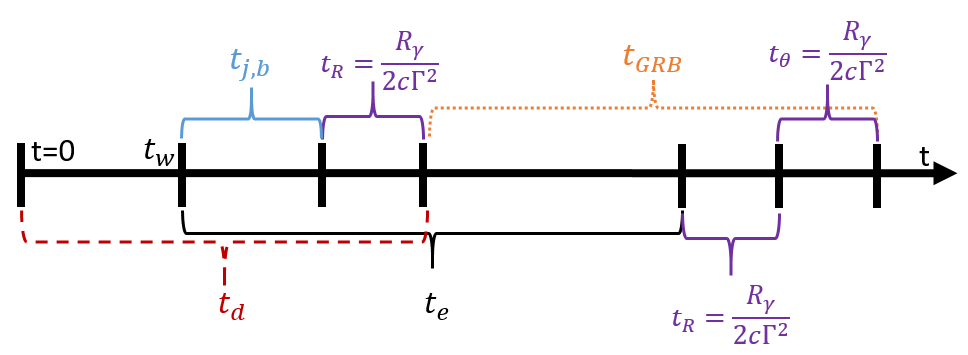}
		\caption{Schematic illustration (not to scale) of the relevant timescales setting the observed duration of the burst, $t_{\rm GRB}$, and the delay time, $t_{\rm d}$, between the burst and a BNS merger occurring at $t=0$. Propagation to the $\gamma$-ray emitting radius $R_{\gamma}$ delays the arrival times of the first and last GRB photons in the same way, thus not contributing to the net GRB duration. 
		}\label{fig:schem}
	\end{figure}

	In addition to the timescales mentioned above ($t_{\rm w}$, $t_{\rm e}$, $t_{\rm j,b}$), there are other important timescales to be considered, which are related to the accompanying observable $\gamma$-ray signal. These are the observer's frame\footnote{For simplicity, we omit the dependence on cosmological redshift in the expressions throughout this paper. The latter can be trivially included by multiplying all observed timescales by a factor of $1+z$.}
	duration of the GRB ($t_{\rm GRB}$), the delay time between the GW and $\gamma$-ray signals ($t_{\rm d}$), the propagation time of the relativistic ejecta (moving at Lorentz factor $\Gamma$) to the $\gamma$-ray emitting radius $R_{\gamma}$, given by $t_{\rm R}\approx R_{\gamma}/2c\Gamma^2$, and the angular timescale associated with the $\gamma$-ray emitting shell ($t_{\rm \theta}\approx R_{\gamma}/2c\Gamma^2$). The latter is the time difference between arrival of photons emitted on-axis to the observer and ones emitted at an angle of $1/\Gamma$ (which due to relativistic beaming is approximately the highest latitude that is visible to the observer).  
	A schematic illustration of the different timescales of the problem is shown in Figure \ref{fig:schem}. 
	
	The delay time between the GW signal and $\gamma$-ray signals is the sum of the following three timescales: the time between the BNS merger and jet launch, the time it takes the jet to break out, and the time it takes the jet to reach the $\gamma$-ray emitting radius,
	\begin{equation}
	\label{eq:td}
	t_{\rm d}=t_{\rm w}+t_{\rm j,b}+t_{\rm R}.
	\end{equation}
	The observed duration of the GRB is given by the difference between the engine and jet breakout times (yielding the amount of time during which a successful jet is passing through $R_{\gamma}$) plus the spreading due to the angular timescale from $R_{\gamma}$:
	\begin{equation}
	\label{eq:tGRB}
	t_{\rm GRB}=t_{\rm e}-t_{\rm j,b}+t_{\rm \theta}.
	\end{equation}
	
	In the following sections, we examine the two limiting regimes (i.e., static ejecta and homologous expanding ejecta). For both regimes, we
	use the observed distribution of GRB durations and luminosities to set limits on $t_{\rm w}$.
	This allows us to determine the validity of the approximations corresponding to both regimes and to place overall limits on $t_{\rm w}$, which hold also for any intermediate regime.

	\subsection{Jet breakout through static ejecta}
	\label{sec:static}
	
	We begin by considering the static ejecta limit.
	In this case, the breakout time of a successful jet is given by the time it takes the jet to overpass the merger ejecta 
	\begin{equation}
	\label{eq:tjb}
	t_{\rm j,b}=t_{\rm w}\frac{\beta_{\rm ej}}{\beta_{\rm h}-\beta_{\rm ej}},
	\end{equation}
	where $\beta_{\rm ej}$ is the velocity of the ejecta and $\beta_{\rm h}$ is the velocity of the jet's head.
	As noted above, self-consistency of this regime requires $t_{\rm j,b}\lesssim t_{\rm w}$ or, equivalently, $\beta_{\rm h}\gtrsim 2\beta_{\rm ej}$ (see equation \ref{eq:tjb}). The velocity of the jet's head is related to the ratio between the jet's isotropic equivalent luminosity, $L_{\rm e}$, and the (isotropic equivalent) mass outflow rate of the ejecta, $\dot{M}_{\rm ej}$, as follows \citep{Marti1994,Matzner2003,Bromberg2011,MB2017}
	\begin{equation}
	\label{eq:betah}
	\beta_{\rm h}=\frac{\beta_{\rm j}+\beta_{\rm ej} \tilde{L}^{-1/2}}{1+\tilde{L}^{-1/2}},
	\end{equation}
	where
	\begin{eqnarray}
	\label{eq:Ltilde}
	& \tilde{L}\!\equiv \!\frac{L_{\rm e}\beta_{\rm ej}}{\dot{M}_{\rm ej}c^2}=\!\\& \!0.14 \frac{f_{\Omega}}{\eta_{\gamma}}\bigg(\frac{L_{\rm GRB}}{10^{52}\mbox{erg  s}^{-1}}\bigg)\bigg(\frac{\beta_{\rm ej}}{0.25}\bigg)\bigg(\frac{10^{-2}M_{\odot}}{M_{\rm ej}}\bigg)\bigg(\frac{t_{\rm w}\!+\!t_{\rm j,b}}{1\mbox{ s}}\bigg).\nonumber
	\end{eqnarray}
	A full derivation of equations  (\ref{eq:betah}), (\ref{eq:Ltilde}) is given in appendix \ref{sec:betah}. In particular we note that even in a mildly relativistic regime, with $\beta_{\rm h}=0.25$, equation \ref{eq:Ltilde} holds to better than a $5\%$ accuracy. In addition, this analytical methodology has been shown by \cite{MB2017} to closely match the results from numerical simulations (with values of $\beta_{\rm h}$ varying by at most $30\%$ between the two).
	To obtain the numerical value in the last expression, we have inserted values typical for a sGRB jet. The conversion between the isotropic $\gamma$-ray luminosity $L_{\rm GRB}$ and the engine power $L_{\rm e}$ can be obtained using the $\gamma$-ray efficiency, $\eta_{\gamma}\equiv L_{\rm GRB}/L_{\rm e}$, which is  $\eta_{\gamma}\approx 0.15$ \citep{Beniamini2015}.
	This conversion implicitly assumes that the degree of jet collimation within the BNS merger ejecta is the same as after the jet has broken out. We discuss the validity of this assumption and its implications in \S \ref{sec:varykilo}.
	We have also assumed that the jet is interacting with the polar component of the merger ejecta. The mass and velocity of the latter can be inferred from the `blue' component of the kilonova \citep{Kasen2017}. It is considered to be associated with the `squeezed' tidal tails, which can be approximated to be roughly isotropically spread up to a polar angle of $\sim \pi/4$ (see e.g. \citealt{Kasen2017}). $\dot{M}_{\rm ej}$ can therefore be approximated by $\dot{M}_{\rm ej}=M_{\rm ej}f_{\Omega}^{-1}/(t_{\rm w}+t_{\rm j,b})$, where $f_{\Omega}=\int_0^{\pi/4}{\rm d}\theta \sin \theta \approx 0.3$ is the solid angle covered by the blue component (assuming a two-sided jet) and $t_{\rm w}+t_{\rm j,b}$ is the time between the BNS merger and the jet breakout; it is used as a proxy of the ejecta expansion time before the jet breakout. 
	
	Equations (\ref{eq:tjb}), (\ref{eq:betah}), and (\ref{eq:Ltilde}) allow us to calculate $t_{\rm w}$ as a function of $L_{\rm GRB}$ and $t_{\rm j,b}$. The time interval between the BNS merger and the launch of the jet, $t_{\rm w}$, is found to increase with increasing values of either $L_{\rm GRB}$ or $t_{\rm j,b}$ as we show below.
	$L_{\rm GRB}$ can be directly constrained from observations for GRBs with redshift determination, while $t_{\rm j,b}$ can be estimated in the following way.
	We first note that $t_{\rm e}\gtrsim t_{\rm j,b}$ is required in order to avoid most of the jet energy to be deposited in the cocoon instead of the GRB jet \citep{RamirezRuiz2002}. In this case, the GRB  duration (in the engine's rest frame) is set by equation (\ref{eq:tGRB})
	(see also Figure \ref{fig:schem}). This relation can be better understood in the following limits:
	\begin{enumerate}
		\item $t_{\theta}\gg t_{\rm j,b}$. In particular, since $t_{\rm GRB}\gtrsim t_{\theta}$  then $t_{\rm GRB}\gg t_{\rm j,b}$ regardless of $t_{\rm e}$. 
		\item  $t_{\theta}\ll t_{\rm j,b}$. In this limit, $t_{\rm GRB}\approx t_{\rm e}-t_{\rm j,b}$. For any distribution of $t_{\rm e}$ that has a non-negligible dispersion (i.e.,   not characterized by a dispersion in $t_{\rm e}$ much smaller than its average, $\sigma_{t_{\rm e}}\ll \bar{t}_{\rm e}$), GRBs with a duration $t_{\rm GRB} \ll t_{\rm j,b}$ would require $t_{\rm e} \approx t_{\rm j,b}$, which would be fine tuned and rare \citep{Bromberg2013}. Specifically, there should be approximately one GRB with $t_{\rm GRB}\approx 0.1t_{\rm j,b}$ ($0.01t_{\rm j,b}$) for every ten (hundred) GRBs with $t_{\rm GRB}\approx t_{\rm j,b}$. 
	\end{enumerate}
	Therefore, independently of the unknown value of $t_{\theta}$, for most GRBs the observed $t_{\rm GRB}$ corresponds to an upper limit on $t_{\rm j,b}$. The derived upper limits are most conservative if one assumes $t_{\theta}\rightarrow 0$. This is because $t_{\theta}$ is an extra component in the GRB duration that is completely independent of the break out. Therefore, a non-zero $t_\theta$ only increases the difference between the GRB duration and the jet breakout time, see equation (\ref{eq:tGRB}). By assuming $t_{\rm GRB}\approx t_{\rm j,b}$ one typically overestimates the true value of $t_{\rm j,b}$ and, in turn,  of $t_{\rm w}$, since the latter increases with $t_{\rm j,b}$. For the purpose of placing upper limits on $t_{\rm w}$ (denoted below as $t_{\rm w,u}$, assuming $t_{\rm GRB}\approx t_{\rm j,b}$ is therefore conservative. As a result, GRBs with short durations and low GRB luminosities place the strongest limits on $t_{\rm w}$ (see \citealt{MB2017} for a similar approach). We demonstrate this point quantitatively in Appendix \ref{app:MC}.
	
\begin{figure}
		\center
		\includegraphics[width=0.5\textwidth]{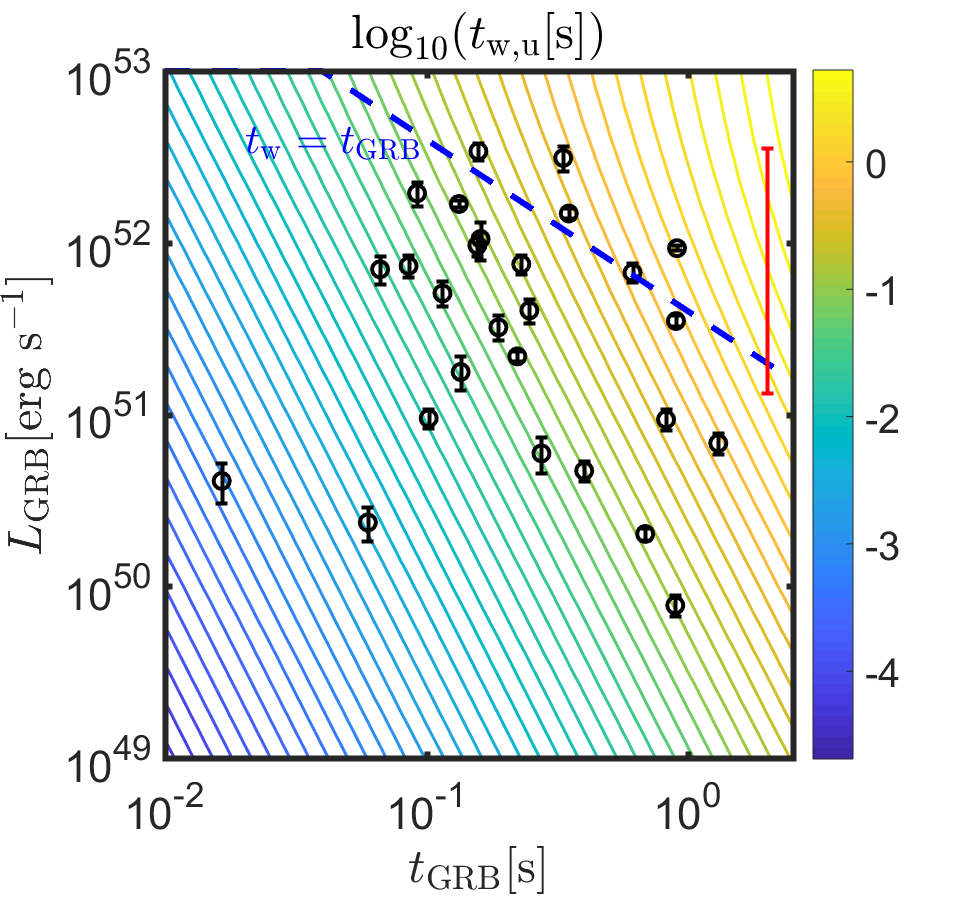}
		\caption{Required (upper limits on the) waiting times (i.e., time intervals between the BNS merger and the launch of the jet -- colored solid lines) needed to account for the observed sGRB durations and luminosities within the static ejecta scenario for $t_{\rm GRB}\approx t_{\rm j,b}$. A dashed blue line marks $t_{\rm w}=t_{\rm j,b}$. Below this line, i.e., for $t_{\rm w}<t_{\rm j,b}$, the assumption of a static medium begins to break down. We overplot the data of observed {\it Swift} short GRBs with known redshift (corrected for the central engine frame) as black circles and the values for the core of GRB 170817 in red (no circle).}\label{fig:twstatic}
	\end{figure}

	Using the assumption that $t_{\rm GRB}\approx t_{\rm j,b}$, as described above, we calculate the value of $t_{\rm w,u}$ for a static medium as a function of $t_{\rm GRB}$ and $L_{\rm GRB}$. These values are plotted in Figure \ref{fig:twstatic}. The distribution of $t_{\rm w,u}$ that is needed to explain the population of the observed 27 GRBs in our sample has a median of $t_{\rm w,u}\approx 0.09$~s and a standard deviation of $\sigma_{\log_{10}(t_{\rm w,u})}=0.7$.
	As shown in Figure \ref{fig:twstatic}, a large fraction of bursts (those below the diagonal dashed line) have no self-consistent solutions with $t_{\rm GRB}\approx t_{\rm j,b}$ and $t_{\rm w}\gtrsim t_{\rm j,b}$ under the static medium scenario. Even if we account for the uncertainty in our model parameters and allow $M_{\rm ej}$ to be reduced by a full order of magnitude from our canonically assumed value, we still cannot find consistent solutions for 8 out of the 27 GRBs ($\sim$30\%). Since this inconsistency cannot be easily resolved by changing the ejecta properties within reasonable bounds, it suggests that, at least in some cases, the homologous expansion limit may be a more realistic assumption than the static medium limit. We shall explore the implications of this approach in the next section. 
	
	\subsection{Jet breakout through homologously expanding ejecta}
	\label{sec:homologous}
	This limit has recently been studied analytically and numerically by D18. Assuming that degree of jet collimation does not change during the jet breakout process, these authors have found that jets are successful when $E_{\rm j}\gtrsim 0.1 E_{\rm ej}$ where $E_{\rm ej}\approx 0.5 M_{\rm ej}\beta_{\rm ej}^2 c^2$ is the kinetic energy of the ejecta and $E_{\rm j}$ denotes the isotropic equivalent energy of the jet\footnote{D18 have used the same notation (i.e., $E_{\rm j}$) to denote the beaming-corrected energy, i.e., $E_{\rm j,D18}=E_{\rm j}\theta_0^2/2$, where $E_{\rm j,D18}$ is the value denoted as $E_{\rm j}$ in D18 and $\theta_0$ is the jet opening angle, hence the difference in the appearance of the equation.}.
	D18 have identified two breakout regimes. For energies in the range $0.1E_{\rm ej} \lesssim E_{\rm j}\lesssim 3 E_{\rm ej}$, jets barely break out and a significant amount of energy is deposited in a cocoon. This regime is dubbed the `late breakout'. For higher energies, $E_{\rm j}\gtrsim 3 E_{\rm ej}$, jets break out easily, and this regime is dubbed `early breakout'. Relating the latter condition to observational properties, we find
	\begin{equation}
	\label{eq:homologous}
	L_{\rm GRB}\!>\! 1.7 \times 10^{51} \eta_{\gamma}\bigg(\frac{\beta_{\rm ej}}{0.25}\bigg)^2\bigg(\frac{M_{\rm ej}}{10^{-2}M_{\odot}}\bigg)\bigg(\frac{1\mbox{ s}}{t_{\rm e}}\bigg)\mbox{ erg s}^{-1},
	\end{equation}
	where we have taken $E_{\rm j}\approx t_{\rm e} L_{\rm e}\approx t_{\rm e} L_{\rm GRB}/ \eta_{\gamma}$. The jet breakout time is given by (D18)
	\begin{eqnarray}
	\label{eq:tjbhom}
	& t_{\rm j,b}=0.3 \, t_{\rm e} \frac{E_{\rm ej}}{E_{\rm j}}=0.15 \, \eta_{\gamma} \frac{M_{\rm ej}(\beta_{\rm ej}c)^2}{L_{\rm GRB}} \quad \mbox { `early'} \nonumber \\
	& t_{\rm j,b}=\frac{9t_{\rm e}}{\sqrt{\frac{10E_{\rm j}}{E_{\rm ej}}\!-\!1}}\!=\frac{9t_{\rm e}}{\sqrt{\frac{20L_{\rm GRB}t_{\rm e}}{\eta_{\gamma}M_{\rm ej}(\beta_{\rm ej}c)^2}\!-\!1}} \quad \mbox { `late'}  
	\end{eqnarray}
	In particular, notice that in the `early breakout'  regime, the jet breakout time becomes independent of the engine duration, and is a function of the jet luminosity only (see also D18 and \citealt{Lyutikov2020}).
	The `early breakout' relation implies that $t_{\rm j,b}<0.1 t_{\rm e}$, since $E_{\rm j}>3E_{\rm ej}$ in this case. Thus, the jet breakout time is sub-dominant in determining the GRB duration, i.e., $t_{\rm GRB}\approx t_{\rm e}-t_{\rm j,b}\approx t_{\rm e}$ (where we have neglected the potential contribution of $t_{\theta}$, see \S \ref{sec:static} for details). 
	
\begin{figure}
		\includegraphics[width=0.5\textwidth]{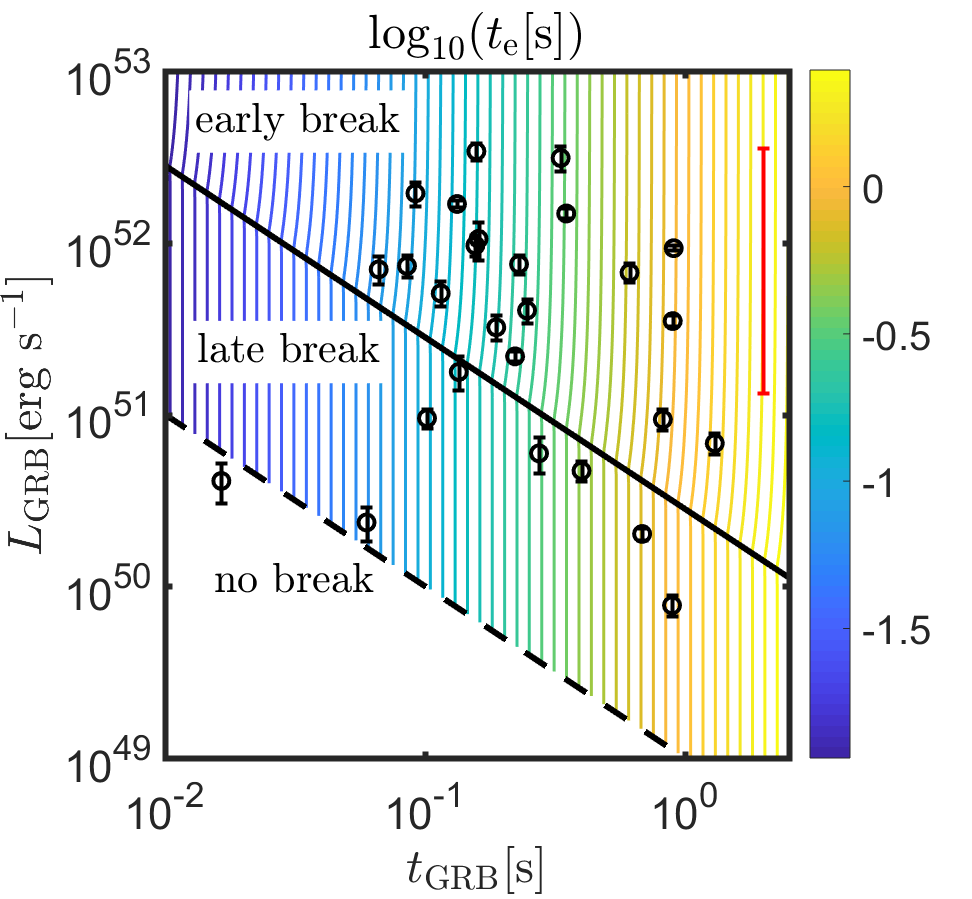}
		\caption{Required engine times (colored solid lines) needed to operate the observed sGRB durations and luminosities for the limit of homologously expanding ejecta. Above the solid line, jets break easily out of the ejecta (i.e., `early breakout'). Between the dashed and solid lines, jets barely break out (i.e., `late breakout'), while below the dashed line, jets no longer break out from the ejecta. 
			Both lines are calculated from equation (\ref{eq:homologous}) using the appropriate expression for $t_{\rm j,b}$ in each regime  (see \S \ref{sec:homologous} for details). Symbols have the same meaning as in Figure \ref{fig:twstatic}.}\label{fig:tehomologous}
	\end{figure}

	We can test the validity of the condition given by equation (\ref{eq:homologous}) by directly comparing with sGRB duration and luminosity data. The comparison of both the early breakout and late breakout conditions to the data is shown in Figure \ref{fig:tehomologous}. The majority of sGRBs (20/27), satisfy the condition given by equation (\ref{eq:homologous}) and reside in the `early breakout' regime. This is consistent with our previous finding that the majority of sGRBs are successful in breaking out of the BNS merger ejecta \citep{Beniamini2019}. A minority of bursts (5/27) nominally reside in the parameter space for late breakouts. However, these may still be consistent with early breakouts given reasonable  changes in the properties of the ejecta (e.g. an ejecta mass lower by a factor of $\sim 4$ or with a velocity lower by a factor $\sim 2$).
	Two bursts (GRB 150101B and GRB 050509B) are close to the limit of jet failure. These are much less likely to have undergone early breakouts, even taking into account variations in the ejecta parameters. We discuss those GRBs in more detail in \S \ref{sec:exception}.

	By construction, in the limit of  homologously expanding ejecta, the waiting time $t_{\rm w}$ must be sufficiently short so that it can be neglected. Thus, $t_{\rm w}$ cannot be directly constrained.  Nonetheless, we know that the waiting times must be shorter than the engine times which, for early breakouts, are comparable to the GRB durations. This condition translates to $t_{\rm w}<0.2$ s, where 0.2~s is the median of the duration distribution of sGRBs in our sample. This upper limit on the waiting time is consistent with our results in \S \ref{sec:static}, but slightly less constraining. In the next section, we derive limits on the waiting time that are applicable to a generic medium.

	\subsection{Jet breakout through a generic medium}
	\label{sec:general}
\begin{figure}
		\center
		\includegraphics[width=0.45\textwidth]{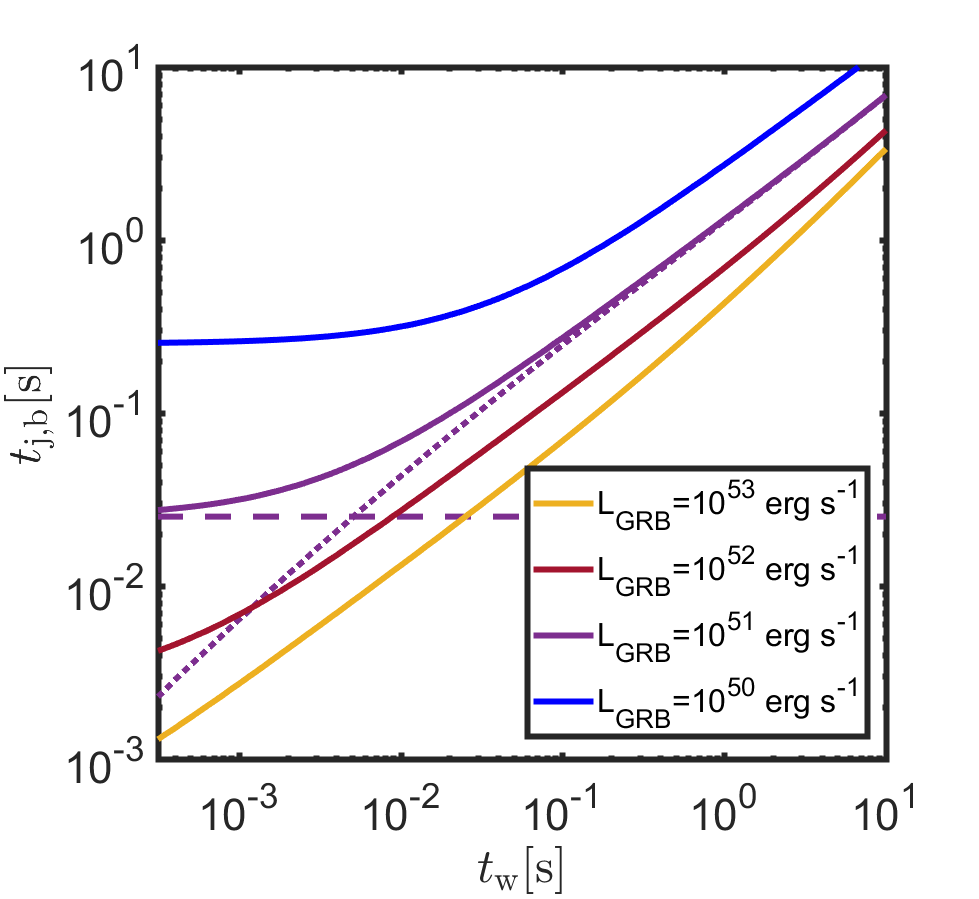}
		\caption{Jet breakout time ($t_{\rm j,b}$) as a function of the time interval between the launch of the BNS merger ejecta and the jet ($t_{\rm w}$) for a generic medium. Results are plotted for $M_{\rm ej}=0.01M_{\odot},\beta_{\rm ej}=0.25, f_{\Omega}=0.3, \eta_{\gamma}=0.15$. Colored curves show the results for different values of the GRB luminosity (see inset legend). For one case ($L_{\rm GRB}=10^{51}$~erg s$^{-1}$), we also show $t_{\rm j,b}(t_{\rm w})$ in the limits of static ejecta (diagonal dotted line) and homologously expanding ejecta (horizontal dashed line). The two curves intersect at $t_{\rm w,c}\approx t_{\rm j,b}/5$.}\label{fig:tjbtw}
	\end{figure}

	Combining the results for the jet breakout time obtained in the  two regimes of static ejecta and homologously expanding ejecta, we can derive limits on the waiting time for a generic medium. For this, let us recall first that the static approximation is formally valid for $t_{\rm w}\gtrsim t_{\rm j,b}$, while the homologous expansion limit is valid for $t_{\rm w}\lesssim t_{\rm j,b}$. Thus, for a generic medium, the jet breakout time must vary continuously between the solutions obtained in these two limits.
	Using equations (\ref{eq:tjb}), (\ref{eq:betah}), (\ref{eq:Ltilde}), (\ref{eq:tjbhom}), the two limits yield identical jet breakout times at a critical waiting time $t_{\rm w,c}\approx t_{\rm j,b}/5$.
	Remarkably $t_{\rm w,c}$ is a function of $t_{\rm j,b}$ only, and it is independent of the other physical parameters. The results are shown in Figure \ref{fig:tjbtw} where we plot $t_{\rm j,b}(t_{\rm w})$ under both approximations. For a generic medium, we can smoothly connect the two regimes by taking
	\begin{equation}
	t_{\rm j,b}=t_{\rm j,b,hom}+t_{\rm j,b,stat},
	\end{equation}
	where $t_{\rm j,b,hom}$, $t_{\rm j,b,stat}$ are the jet breakout times in the homologous expansion and static ejecta limits, respectively.

\begin{figure}
		\center
		\includegraphics[width=0.5\textwidth]{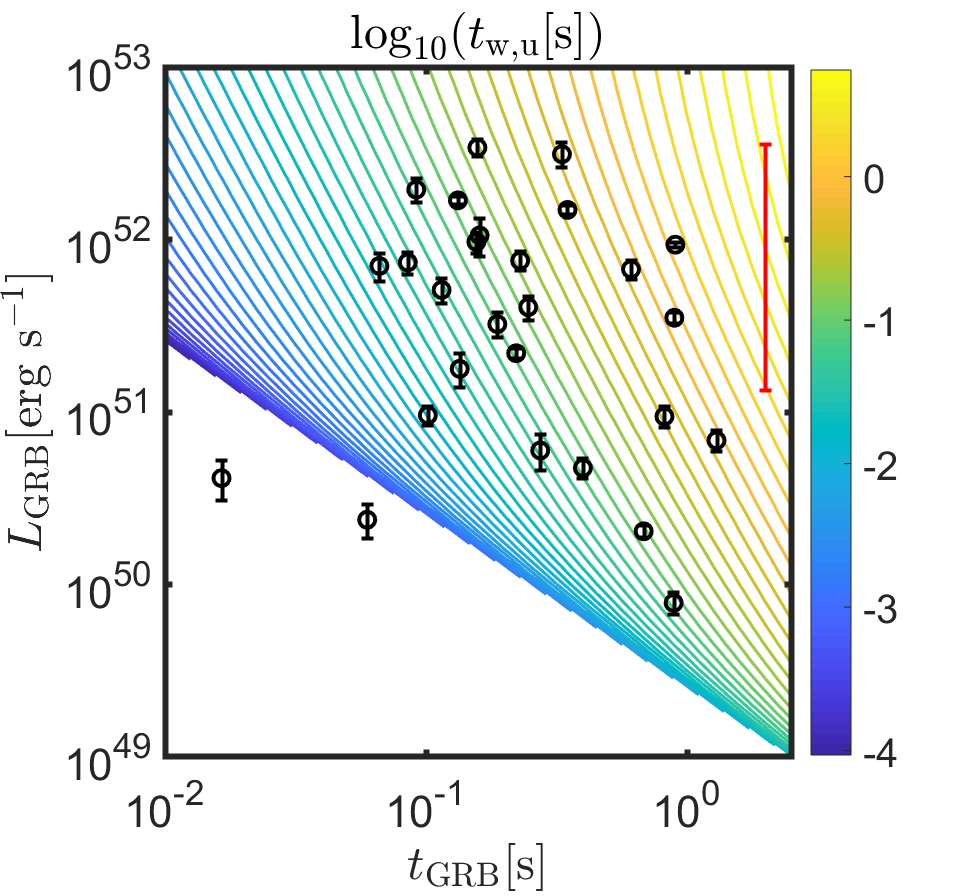}
		\caption{Same as Figure \ref{fig:twstatic} but for a generic medium and for $t_{\rm GRB}\approx t_{\rm j,b}$. Using a generic medium introduces a lower limit on the jet breakout time (corresponding to the homologously expanding medium regime, see Figure \ref{fig:tjbtw}) and restricts the allowed parameter space compared to Figure \ref{fig:twstatic}.}\label{fig:twcombined}
	\end{figure}

	Using the expression for $t_{\rm j,b}$ above we can now obtain upper limits on $t_{\rm w}$ (i.e. $t_{\rm w,u}$) for a generic medium in a similar way to the one  outlined in \S \ref{sec:static}. For a given GRB luminosity, there is a lower limit on the jet breakout time (corresponding to breakout from a homologously expanding medium, see Figure \ref{fig:tjbtw}). This limit decreases with increasing luminosity.
	The result, shown in Figure \ref{fig:twcombined}, is that for sufficiently short durations and / or low luminosities no self-consistent solutions exist (without changing the ejecta properties and / or the $\gamma$-ray efficiency). Within this generic medium scenario, we find again that for the same 2 out of 27 GRBs in our sample, GRB 050509 and GRB 150101B, no solutions ae available (consistent with our findings in \S \ref{sec:homologous}). We return to discuss those bursts in more detail in \S \ref{sec:exception}. Since breakout is more difficult (i.e., $t_{\rm j,b}$ becomes longer) when $t_{\rm w}$ becomes longer, these two bursts likely correspond to shorter waiting times than found for the rest of the sGRB population. Nonetheless, to be more conservative, we ignore these bursts when calculating the upper limits on $t_{\rm w}$ below.
	For the majority (25/27) of GRBs however, we can self-consistently treat the jet breakout. This leads to upper limits on the waiting times. The median upper limit obtained for those bursts is $t_{\rm w,u}\lesssim 0.1$ s (for the dependence of this result on kilonova ejecta properties, see \S \ref{sec:varykilo}).
	
	\section{discussion}
	\label{sec:discuss}
	We have considered the breakout of GRB jets from the BNS merger ejecta. The observed properties of cosmological GRBs, as well as constraints on the merger ejecta from the kilonova counterpart to GW170817 allow us to put limits on the time intervals between the launching of the BNS merger ejecta components and the launching of the relativistic jet. For a generic description of the BNS merger ejecta (that smoothly connects the regimes of static and homologously expanding ejecta), we derive a rough upper limit of $t_{\rm w}\lesssim0.1$~s. 
	Furthermore, we argue that for a fraction  of sGRBs (at least $\sim$30\%) the assumption of static ejecta, through which the jet punches, is inconsistent with their observed luminosities and durations, even if the polar component of the ejecta mass is a factor of ten lower than the one estimated for the kilonova accompanying GW~170817. At the same time, we find that the assumption of homologously expanding ejecta (corresponding to the limit $t_{\rm w}\rightarrow 0$) is consistent with the observed properties ($t_{\rm GRB}$, $L_{\rm GRB}$) of our sGRB sample. The derived upper limit on the waiting time has several interesting implications which we discuss below.
	
	\subsection{Exceptional GRBs}
	\label{sec:exception}
	As mentioned in \S \ref{sec:homologous}-\S\ref{sec:general} two bursts (GRB 150101B and GRB 050509B) are close to  jet failure, even when considering the limit of negligible waiting time ($t_{\rm w}\to 0$), for which breakout becomes easiest. These are much less likely to have undergone early breakouts, even taking into account variations in the ejecta parameters.
	
	One possibility is that the prompt GRB in those cases represents a jet that failed to break through the merger ejecta. In such a scenario a $\gamma$-ray signal may still result due to the shock breakout from the cocoon created by the jet-merger ejecta interaction. In the case of GRB~150101B, \cite{Burns2018} have found evidence for a short hard spike followed by a soft tail, similar to the GRB counterpart of GW~170817, suggesting shock breakout as a possible explanation for the observed $\gamma$-rays. Furthermore, GRB~150101B exhibited a bright optical counterpart consistent with a blue kilonova \citep{Troja2018} with $M_{\rm ej}>0.02M_{\odot}$ and $\beta_{\rm ej}>0.15$. These values are consistent with a late shock breakout as an explanation for this burst. Nevertheless, it is important to note that GRB 150101B does not appear to satisfy the closure relationship between the energy, duration and temperature of shock breakout flares \citep{NakarSari2010}. Assuming this relation, we would expect the peak energy of GRB 150101B to be $\approx 2$~MeV, whereas the observed peak energy is of the order of $550$~keV for the initial spike and much lower, $\sim 23$~keV for the soft tail \citep{Burns2018}. In the case of GRB 050509, the breakout estimate for the temperature yields $k_B T\approx 1.5$~MeV. Since $\nu F_{\nu}$ is seen to be rising within the observed {\it Swift} $15-150$~keV band \citep{Bloom2006}, only a lower limit on the peak energy can be obtained and the possibility of a shock breakout association satisfying the closure relation cannot be ruled out. 
	A major shortcoming of the shock-breakout interpretation for both these bursts, however, is that their $\gamma$-ray luminosities, $\sim 3\times 10^{50}\mbox{ erg s}^{-1}$, are large compared to expectations from shock breakout \citep{NakarSari2010}. Since the shock breakout mechanism releases only a very small fraction of the total energy (see \S \ref{sec:shockbreak}), this interpretation would require much larger engine luminosities to work. 
	This requirement, in turn, would make it much less likely for the jets of those GRBs to have failed to break through the ejecta in the first place (more powerful jets are easier to break out). 
	
	The above discussion makes us consider an alternative (and easier to accommodate) scenario for both GRBs. According to this scenario, $\gamma$-rays are still produced within a successful relativistic jet, as in regular cosmological sGRBs, but the $\gamma$-ray efficiency is much lower than typically used (i.e., $\eta_\gamma \ll 0.1$).
	One natural way for this to happen, is if these GRBs are viewed slightly off-axis from the cores of their jets (see e.g. \citealt{Beniamini2019,Mandhai2019,Bartos2019,Dichiara2020}).

	\subsection{Variation of jet and kilonova ejecta properties}
	\label{sec:varykilo}
\begin{figure}
		\center
		\includegraphics[width=0.5\textwidth]{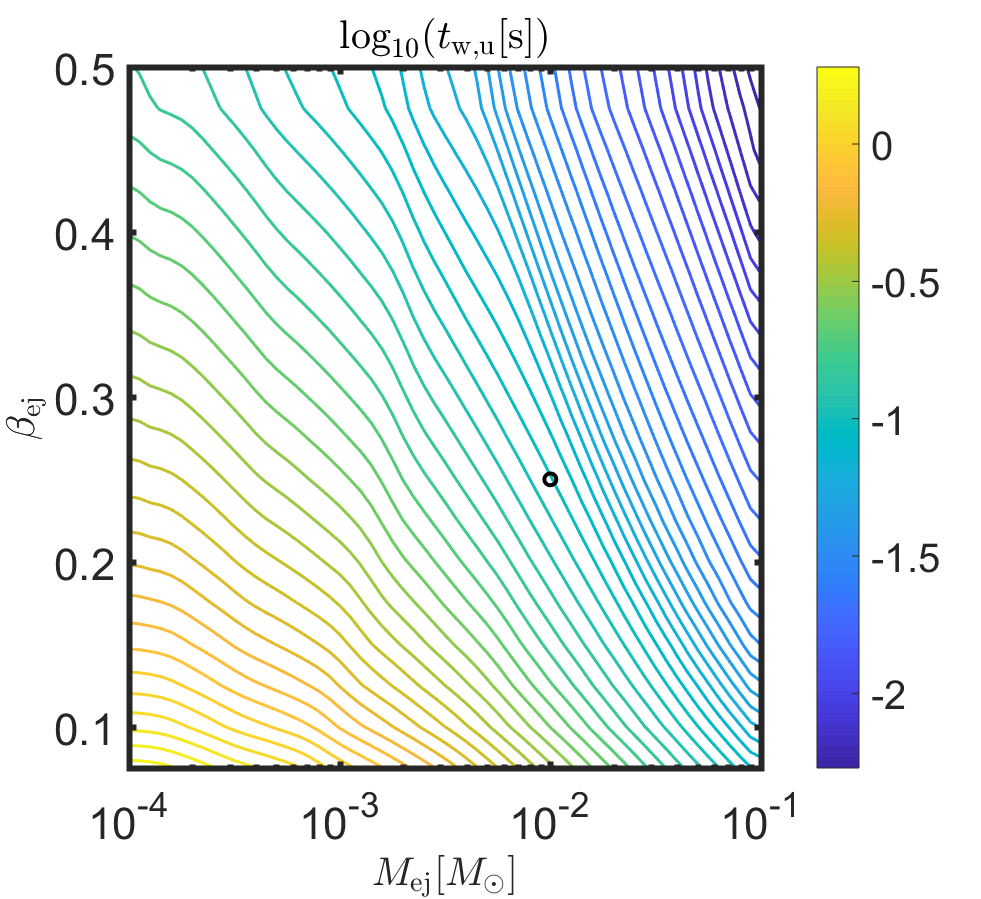}
		\caption{The dependence of the median upper limit on the waiting time on the ejecta mass and velocity corresponding to the blue component of the kilonova. The estimates of those values for GW 170817, that are used elsewhere in this paper, are marked by a circle.}\label{fig:kilovary}
	\end{figure}

	We have argued that taking the ejecta properties to be similar to those inferred from the kilonova accompanying GW~170817, the majority of observed short GRBs would not have been able to break out through a static ejecta. An alternative option, is that there is a very wide variation in the BNS ejecta properties of different BNS mergers. We caution the reader that if indeed the ejecta mass varies very widely between different events, this would tend to increase our upper limits on the waiting times discussed above. The required level of ejecta mass variation under the static ejecta interpretation, however, would be very large. For example, lowering the ejecta mass by roughly two orders of magnitude as compared with the inferred values for GRB~170817 is required in order to enable a successful breakout of all GRBs in our sample under this interpretation. The general dependence of our limit on the waiting time on the ejecta mass and velocity is depicted in Figure \ref{fig:kilovary}. Future observations of kilonovae accompanying GW events from nearby BNS mergers would enable to directly test the validity of this possibility.
	
	Another potential caveat regards the possibility of self collimation. For long GRB jets, that are propagating through the envelope of the collapsed star, self-collimation is expected to make the effective opening angle of the jet, as it is passing through the stellar envelope, narrower than its final opening angle after breakout \citep{Bromberg2011}. For short GRBs, the smaller amount of ejecta mass and the expanding nature of the ejecta, suggest that self-collimation plays a lesser role, especially in the homologous expansion limit \citep{Duffell2018,GNR2019,Hamidani2020}. If however the jet were indeed significantly narrower during their propagation through the BNS merger ejecta, this would result in an effective increase of the isotropic equivalent engine luminosity, and in turn, an increase in $\tilde{L}$ and an increase in our upper limit on $t_{\rm w}$. 
	To test the importance of self-collimation, we have considered a situation in which the effective area of the jet decreases by a full order of magnitude within the merger ejecta (i.e. increasing $\tilde{L}$ by a similar amount). Our limit on the waiting time changes in that case from $t_{\rm w}<0.1$ s to $t_{\rm w}<0.25$ s.
	
	Expanding further on the previous point, one may question how well does the analytical treatment adopted in this work represent the true physical situation. Comparison of analytical and numerical studies \citep{Mizuta2013,Harrison2018,Hamidani2020}, suggest that while the analytic results match rather well the numerical ones they tend to slightly overestimate $\tilde{L}$. This will correspond to a slight decrease in our upper limit on $t_{\rm w}$, making our analytical treatment conservative from this perspective.
	
	\subsection{Engine duration distribution}
	\label{sec:engineduration}
	In either the static or the homologous expansion case, very luminous short GRBs have jet breakout times that are much shorter than the GRB duration. The result is that the GRB duration is almost the same as the engine duration and implies that the duration distribution of luminous bursts directly maps the  distribution of engine durations. At the moment, the numbers of such events are rather low, due to the sparsity of sGRBs with redshift determination and the intrinsic rarity of the most luminous bursts. Increasing this sample in the future, would enable us to glean critical information regarding the nature of the central engine.
	
	Since the difference between the observed and source frame durations of short GRBs is typically less than a factor of two, the duration distribution of short GRBs can be studied even for GRBs with no redshifts. This has the advantage of increasing the data-set significantly, but at the cost of removing  information about the intrinsic luminosity. \cite{Moharana2017}, using a large sGRB sample (with and without $z$ measurements) within the framework of \cite{Bromberg2012}, found possible evidence for a plateau in the duration distribution ($dN/dt_{\rm GRB}$) at $t_{\rm j,b}\approx 0.4$~s, followed by power-law-like distribution at longer durations, $dN/dt_{\rm GRB}\propto t^{-1.4}$. Such a plateau is expected if the jet breakout time of the most commonly observed sGRBs is also $\approx 0.4$~s. This interpretation is consistent with $t_{\rm w}\lesssim0.1$~s, if the most commonly observed sGRBs have characteristic luminosities of $L_{\rm ch}\sim 3\times 10^{50}\mbox{erg s}^{-1}$ (see Figure \ref{fig:tjbtw}). This is consistent with the results of \cite{WP2015},  who showed that the number of sGRBs is dominated by the low end of the observed luminosity function (i.e., $L_{\rm min}\approx 5\times 10^{49}\mbox{erg s}^{-1}$). The proximity of $L_{\rm ch}$ to $L_{\rm min}$ and, in particular, the fact that $L_{\rm ch}$ is orders of magnitude smaller than the characteristic break of the luminosity function, $L_*\approx 2\times 10^{52} \mbox{erg s}^{-1}$, suggest that it is unlikely that $L_{\rm ch}$ plays any significant role in shaping the sGRB luminosity function (see \citealt{Beniamini2019} for details). This is consistent with the conclusions of \cite{Beniamini2019}, namely that the fraction of failed jets cannot explain the broken-power law nature of the sGRB luminosity function, opposite to the case of long GRB jets \citep{Petropoulou2017}. 
	
	\subsection{Extension of the analysis to GRBs with no redshift determination}
	As pointed out in \S \ref{sec:engineduration}, considering GRBs with undetermined redshift has the advantage of significantly increasing the sample size, but at the cost of leaving the luminosity highly uncertain. For this reason, we consider now the 14-yr {\it Swift} sGRB sample that consists of 119 sGRBs without redshift determination only as a consistency check to the main results presented in \S \ref{sec:general}. For this purpose, we make the simplifying assumption that all {\it Swift} GRBs without redshift determination originate from the same redshift $z_0$. 
	As test values we consider $z_0=0.55$, which is the median redshift of GRBs in our sample and $z_0=0.9$, which is roughly the peak of the sGRB redshift distribution found by \cite{WP2015}. The limits on $t_{\rm w}$ can then be obtained in a similar way to that described in \S \ref{sec:general}. The results are $t_{\rm w,u}\lesssim 0.09$ s ($t_{\rm w,u}\lesssim0.12$ s) for $z_0=0.55$ ($z_0=0.9$).
	The proximity of these values to the upper limits derived from the sample of bursts with redshift suggests that our results can be reasonably extended to the general sGRB population.
	
	\subsection{Shock breakout energy}
	\label{sec:shockbreak}
	The energy released during the breakout phase is limited by the thermal energy stored in the cocoon, $E_{\rm Th}$. The latter is limited by the (collimation corrected) energy deposited in the cocoon before the moment of breakout, $E_{\rm Th}\lesssim \frac{\theta_0^2}{2}L_{\rm e} t_{\rm j,b}$. Using the relation $t_{\rm w}( t_{\rm j,b}, L_{\rm GRB})$ derived in equations (\ref{eq:tjb}), (\ref{eq:betah}), (\ref{eq:Ltilde}), (\ref{eq:homologous}), (\ref{eq:tjbhom}), we find that $E_{\rm Th}\lesssim 4\times 10^{49}$~erg for $t_{\rm w}\lesssim 0.1$~s and $L_{\rm e}=10^{53}\mbox{ erg s}^{-1}\approx L_*/\eta_{\gamma}$. If the engine power is reduced or the waiting time is shorter the upper limits on $E_{\rm Th}$ would become more constraining. As a comparison, in the homologous case, the thermal energy of the cocoon is (for both the early and late breakout scenarios) $E_{\rm Th} \lesssim 5 \times 10^{47} (\theta_0/0.1)^2 \mbox{ erg}$ (see \citealt{Beniamini2019} for details).
	Furthermore, since initially the cocoon is highly optically thick \citep{NakarSari2010}, only a small fraction of this energy is expected to be released as prompt $\gamma$-rays during the breakout phase. Overall, we expect the quasi-isotropic shock breakout signal accompanying sGRBs to be typically rather weak.
	
	\subsection{The delay time of GRB~170817}
	\label{sec:Delaytime}
	The observed delay of $t_{\rm d}\sim 1.7$ s between the GW and the $\gamma$-ray signal from GRB 170817 \citep{GW170817} can be expressed as the sum of $t_{\rm w}$, $t_{\rm j,b}$, and $t_{\rm R}$ (see equation \ref{eq:td} and Figure \ref{fig:schem}). A natural question arises then: Which of the three timescales dominates the observed $t_{\rm d}$? Since the first two timescales are a function of the jet luminosity, the answer depends critically on the luminosity of GRB 170817 along its core. 
	
\begin{figure}
		\center
		\includegraphics[width=0.5\textwidth]{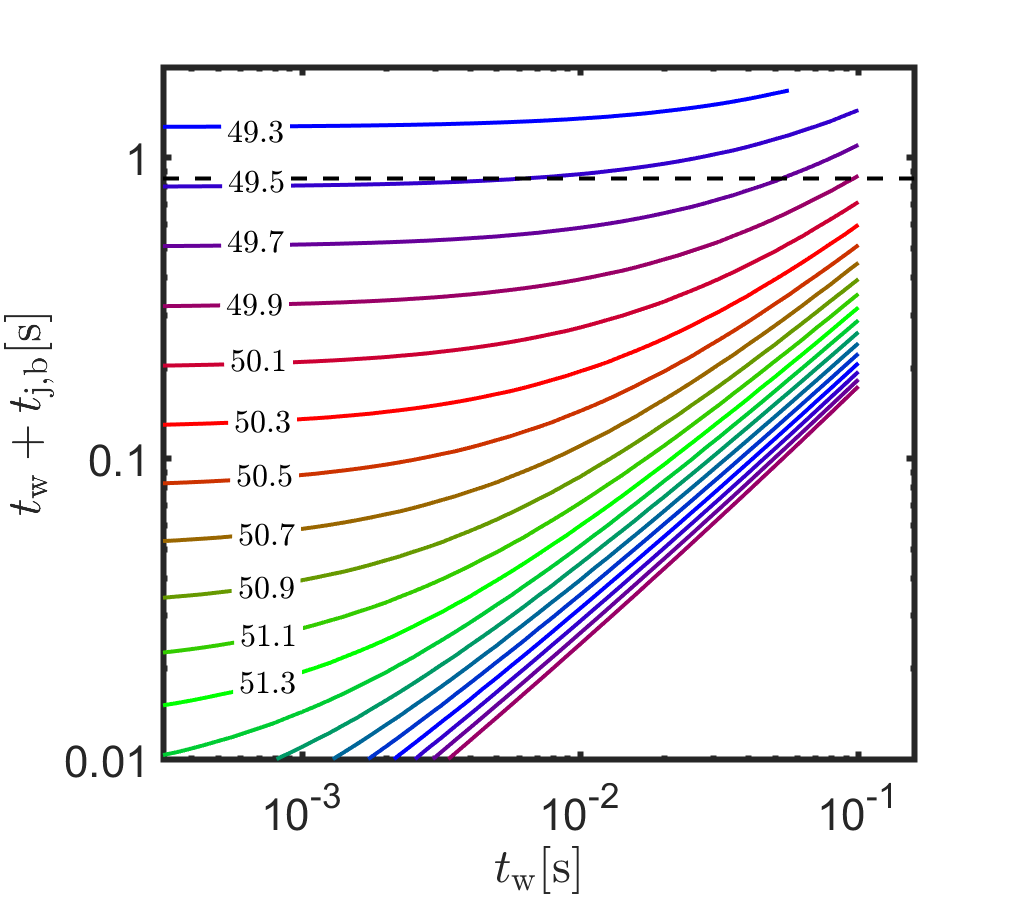}
		\caption{GRB luminosity along the core of the jet and waiting times that correspond to delay times between GW and $\gamma$-rays that are consistent with observations in GRB 170817. The curves correspond to combinations of these properties that ensure $t_{\rm w}<0.1$ s (as found in this paper) and $t_{\rm w}+t_{\rm j,b}<1.7$ s (as required by the observed delay). From top to bottom contours depict jet core isotropic equivalent GRB luminosities with equal logarithmic intervals (labels indicate $\log_{10}(L_{\rm GRB}$)). A dashed horizontal line marks $t_{\rm w}+t_{\rm j,b}=t_{\rm d}/2$. Afterglow fitting of GRB~170817 strongly favours $L_{\rm GRB}\gtrsim 1.3\times 10^{51} \mbox{erg s}^{-1}$, which is the region below the solid green line, implying that $t_{\rm w}+t_{\rm j,b}<t_{\rm d}/2$.}\label{fig:tdelay}
	\end{figure}

	In Figure \ref{fig:tdelay} we plot the allowed parameter space given by the requirements $t_{\rm j,b}+t_{\rm w}<1.7$~s and $t_{\rm w}<0.1$ s. We caution the reader that the latter constraint is found from a statistical analysis of the entire sGRB sample. It is of course possible that for any specific event, the waiting time may be longer. Indeed several groups have demonstrated that they can reproduce the observed signatures of GRB 170817 with numerical simulations involving longer ($\sim 1$ sec) delays \citep{Mooley2018,Xie2018}.
	In addition, \cite{GNR2019} have shown that such longer delays could be favourable for explaining the observed properties of the associated kilonova emission in that event. Nonetheless, we expect the analysis outlined here to be representative of future GW detected sGRBs. \cite{Mooley2018} have demonstrated that GRB 170817 involved a powerful jet that broke out of the merger ejecta.
	Given the energy at the core of GRB 170817 inferred from afterglow fitting \citep{Troja2019}, its GRB isotropic equivalent luminosity (along its core) is estimated to be $L_{\rm GRB}\gtrsim 1.3\times 10^{51} \mbox{erg s}^{-1}$ (region below the solid green line in Figure~\ref{fig:tdelay}), clearly in contradiction with $t_{\rm j,b}+t_{\rm w}>t_{\rm d}/2$ (region {\it above} the horizontal dashed line in Figure~\ref{fig:tdelay}).
	This implies that $t_{\rm j,b}+t_{\rm w}<t_{\rm d}/2$ for GRB~170817, which translates to $t_{\rm d}\approx t_{\rm R}\approx R_{\gamma}/2c\Gamma^2$. It is worth noting that due to angular spreading, this situation can naturally lead to a pulse duration which is also of the order of $t_{\theta}\approx R_{\gamma}/2c\Gamma^2$. This is realized in prompt emission models for which that other timescales involved in the prompt GRB phase, such as the cooling time, and engine variation timescale are shorter or equal to $t_{\rm \theta}$. Models of this kind include the internal shocks model \citep{SP1997,DM1998} and several magnetic reconnection based models where dissipation takes place far from the central engine (e.g., \citealt{KN2009,Zhang2011,BG2016,BG2017,BBDG2018}). Interestingly, the duration of GRB~170817 was of the same order as the time delay. If future events continue to show a similar trend, this will be a strong indication in favour of $R_{\gamma}/2c\Gamma^2$ dominating the observed time delay (see \citealt{Zhang2019}, for a detailed discussion on this point). 
	
	As opposed to regular cosmological GRBs, which are observed on-axis, GRB 170817 was observed off-axis \citep{Mooley2018}. As a result, the propagation and angular timescales ($\propto R_{\gamma}/\Gamma^2$) are likely to be larger than for on-axis GRBs.
	The extent of this effect depends on the nature of the mechanism producing the prompt emission. For example, in photospheric models, $R_{\gamma}\propto L_{\rm e}\Gamma^{-3}$ leading to $R_{\gamma}/\Gamma^2\propto L_{\rm e}\Gamma^{-5}$. For typical expected profiles of power and Lorentz factor in the jet (e.g. \citealt{Kathirgamaraju2018,BN2019}), the latter is an increasing function of polar angle, suggesting  longer propagation and angular timescales for GRBs seen off-axis.
	
	Regardless of the specific prompt emission model, a measurement of the delay, $t_{\rm d}$, leads to a lower limit on the Lorentz factor of the prompt producing material due to the following argument.
	Radiation is decoupled from the jet material when the Thomson optical depth $\tau_{\rm T}$ becomes of order unity. This happens at the so-called photospheric radius, which can be determined by setting $\tau_{\rm T}=1$ as: $R_{\rm ph}\simeq 10^{14} L_{\rm e,47}\Gamma^{-3}$~cm, where  $L_{\rm e}=10^{47} \, L_{\rm e,47}$~erg~s$^{-1}$ is the jet's isotropic-equivalent power of the material that dominates the emission towards the observer (which in the case of GRB 170817, was outside of the jet core, e.g. \citealt{Finstad2018}) and $\Gamma$ is the Lorentz factor of the same material (see, e.g., \citealt{Giannios2012}). Using the observed time delay $t_{\rm d}$ and noting that $R_{\gamma} \gtrsim R_{\rm ph}$, we can derive a lower limit on the bulk Lorentz factor of said material that is independent of the prompt emission model: $\Gamma\gtrsim 4 \, L_{\rm e, 47}^{1/5} \left(t_{\rm d}/1.7 {\rm s}\right)^{-1/5}$. 
	Note that the exact numeric value is not so sensitive to the jet power of the material dominating the observed emission. Furthermore, this material need not necessarily lie directly along the line of sight. The argument outlined here would hold for material from an intermediate angle $\theta_0<\theta<\theta_{\rm obs}$, such that it is Doppler boosted towards the observer, i.e. $(\theta_{\rm obs}-\theta)\Gamma<1$. 
	
	\subsection{Nature of central engine}
	In any GRB central engine, the jet breakout time $t_{\rm j,b}(t_{\rm w},L_{\rm e})$  must be large enough that by the time the jet breaks out and starts emitting, its bulk Lorentz factor is sufficiently high to avoid the compactness problem (i.e., to avoid a very large optical depth of the emitting material, see definition of $R_{\rm ph}$ in \S \ref{sec:Delaytime}). 
	
	For magnetar central engines, where the flow is initially heavily baryon loaded, this is not easily achieved.
	Since the power released by the magnetar, $\dot{E}$, typically decreases slower in time as compared to the mass outflow rate, $\dot{M}$ \citep{Metzger2011}, there is generally a minimum time, $t_{\rm d,mag}$, before the energy per baryon at the base of the jet ($\eta \propto \dot{E}/\dot{M}$) becomes sufficiently high (i.e., $\eta\gtrsim 100$) to power an ultra-relativistic GRB \citep{BGM2017}.
	For typical parameters of the magnetar model, $t_{\rm d,mag}\gtrsim 3$~s. The main parameter affecting this timescale is the magnetar's magnetic field, $B$.  Only for an extreme value of $B\sim 3\times 10^{16}$ G, one finds $t_{\rm d,mag}\approx 0.2$~s.
	As shown in Figure \ref{fig:tjbtw}, this is still too high given the expected values of $t_{\rm j,b}$ with a waiting time of $t_{\rm w}\lesssim0.1$ s and a (rather common) $L_{\rm GRB}\gtrsim 3\times 10^{51}\mbox{ erg s}^{-1}$.
	Fall-back accretion onto the magnetar, may alter the timescale $t_{\rm d,mag}$, but given the high accretion rates expected for binary neutron star mergers, this effect tends to reduce the initial energy per baryon and therefore increase $t_{\rm d,mag}$ even more \citep{MBG2018}. As a result we conclude that magnetar central engines are severely challenged as possible engines of short GRBs.

	In the context of black hole central engines, a waiting time of $t_{\rm w}\lesssim 0.1$ s, suggests a relatively prompt collapse of a neutron star to a black hole.
	This suggests that the remnant mass from the binary neutron star merger should be massive enough to form at least a hypermassive neutron star (a short lived neutron star supported by differential rotation), see also \cite{MB2014,MB2017}. 
	The collapse time to a black hole depends on the tidal deformability parameter and on the equation of state of the neutron star \citep{FH2008,Favata2014}. For example, \cite{Radice2018} have shown that for GW~170817, this time is indeed expected to be 1-10~ms, consistent with our limit on $t_{\rm w}$.
	
	Physically, the waiting time can be composed of the sum of time it takes to form the central black hole and the time it takes it to accrete a significant amount of mass. Our limit on the waiting time, therefore limits also the accretion time, $t_{\rm acc}<t_{\rm w,u}$. Recent simulations of magnetically launched  jets from neutron star merger accretion discs \citep{Christie2019} find that the jet power peaks within $\lesssim 0.05$ s. This is consistent with the results found in this paper.
	Useful intuition on the accretion timescale can be obtained from an $\alpha$-viscosity thick disk model for the accretion. This limit has been applied to different neutron star merger simulations by \cite{Fryer2015} to constrain the outcome of the merger as a function of e.g. the individual neutron star masses and the equation of state. Using $t_{\rm acc}=2\pi r^{3/2}/(\alpha \sqrt{G M_{\rm enc}})$ (where we assume that immediately after the merger there is a thick disk with radius $r$, enclosing a mass $M_{\rm enc}$) and requiring $t_{\rm acc}<t_{\rm w,u}$ we can constrain the $\alpha$ viscosity parameter of the disk,
	\begin{equation}
	\label{eq:tacc}
	\alpha\gtrsim0.01 \bigg(\frac{r}{2\times 10^6\mbox{ cm}}\bigg)^{3/2} \bigg(\frac{0.1\mbox{ s}}{t_{\rm w,u}}\bigg) \bigg(\frac{2.6M_{\odot}}{M_{\rm enc}}\bigg)^{1/2} .
	\end{equation}
	In the last expression we have considered a radius for the disk immediately after merger which is of the order of two neutron star radii (in general this can be considered as a lower limit). We also took the enclosed mass to be approximately the minimum total mass required for producing a hyper-massive neutron star \citep{Baumgarte2000ApJ...528L..29B} that would quickly collapse to a black hole (a significantly smaller mass would correspond to a long-lived neutron star in contrast with our limits on the waiting time, as discussed above, while a significantly higher mass would lead to a prompt collapse and would result in little amount of mass in the disk that would be available to power the following GRB). This mass is also consistent with the total mass of known Galactic binary neutron stars (e.g. \citealt{BP2016}).
	Our limits on $\alpha$ from equation \ref{eq:tacc} are consistent with the considerations of \cite{Fryer2015}.
	
	\section{Conclusions}
	\label{sec:conclude}
	
	We have revisited the conditions for breakout of a sGRB jet from the BNS merger ejecta.
	Using published results from analytical and numerical works on this topic, which apply either to the case of static merger ejecta or homologously expanding ejecta (e.g., \citealp{Begelman1989,Marti1994,Duffell2018}), we derive the conditions of successful jet breakout for a generic medium that smoothly connects these limiting cases. 
	Using the {\it Swift}-BAT sample of sGRBs with measured redshift, we derive limits on the waiting time, i.e., the time interval between the BNS merger and the launching of the jet (assuming that the BNS merger ejecta is launched immediately after the BNS merger).
	
	For all the cases we examined (i.e., static ejecta, homologously expanding ejecta, and generic medium), we set an upper limit of $\sim 0.1$~s on the waiting time.  Decreasing the ejecta mass (velocity) by a factor of ten (two), increases the upper limit on $t_{\rm w}$ by a factor of $\sim 4$ (2). Our results on the waiting time can be also extended to the complete {\it Swift} sample of sGRBs with no redshift determination. 
	Our upper limit on the waiting time is consistent with previous results (e.g., \citealp{MB2014,MB2017}) obtained with a smaller sample and in the limit of static merger ejecta. We also show that this typically adopted limit for the BNS ejecta is inconsistent with at least $\sim 30\%$ of our sGRB sample.  
	
	Although the analytical treatment adopted in this paper is approximated, and does not take into account some of the finer details of jet propagation observed in numerical simulations, such as collimation shocks, this treatment is in good agreement with several numerical studies \citep{MB2017,Duffell2018,Hamidani2020}. We stress that our overall result is rather intuitive. Given that sGRBs typically last $\sim 0.2$ s and that the rate of BNS mergers and successful sGRB jets are similar \citep{Beniamini2019}, it is unlikely that the characteristic breakout and waiting timescales could be much longer than the sGRB durations, as this would require a fine tuning between the engine activity time and these timescales.
	
	The limit on the interval between the BNS merger and the launching of the jet has profound consequences for the origin of $\gamma$-ray emission (i.e., cocoon shock breakout versus jet) and the nature of the sGRB central engine (i.e., magnetar versus black hole). It restricts the amount of thermal energy stored in the cocoon (e.g., $E_{\rm Th}\lesssim 4\times 10^{49}\mbox{ erg}$ for $L_{\rm e}=10^{53}\mbox{ erg s}^{-1}$), suggesting that the shock breakout signal accompanying sGRBs is expected to be rather weak. It also suggests that central engines of sGRBs are unlikely to be milli-second magnetars (i.e., with $B\lesssim 3\times 10^{16}$~G), since the time interval of $\lesssim 0.1$~s is too short to produce a jet with sufficiently high energy-per-baryon at its base to allow its bulk acceleration to ultra-relativistic speeds.
	Our results are therefore in favor of a relatively prompt collapse  (i.e., within $<100$~ms) of a neutron star to a black hole.  
	
	In the context of GRB~170817 our work places strong constraints on the physical origin of the observed $\gamma$-ray signals, assuming that the statistical limit on $t_{\rm w}$ found in this work, applies also for this specific GRB. We find that the observed delay between the GW and the $\gamma$-ray signal is dominated by the time it takes the jet to reach the location at which it will radiate (see also \citealp{Zhang2019}). The consequence of this interpretation is that the $\gamma$-ray duration may naturally (depending on the prompt emission model, see \S \ref{sec:Delaytime}) be of the same order of the observed delay, which is the case for GRB~170817. Future observations would indicate if this is the case for other bursts. This could provide a much needed independent test for comparing between the many prompt emission models that still remain viable at this point.

	\section*{Acknowledgments}
	The research of PB was funded by the Gordon and Betty Moore Foundation through Grant GBMF5076. RBD and DG acknowledge support from the National Science Foundation under Grants 1816694 and 1816136. MP acknowledges support from the Lyman Jr.~Spitzer Postdoctoral Fellowship and the Fermi Guest Investigator Program Cycle 12, grant 80NSSC18K1745. DG acknowledges support from the NASA grant NNX17AG21G and the Fermi Guest  Investigator Program Cycle 12, grant 80NSSC19K1506. 
	

	\appendix
	\section{Estimating the jet head's velocity}
	\label{sec:betah}
	The jet head velocity can be found by the requirement of pressure balance between the shocked jet head material and the shocked ambient medium (see, e.g. \citealt{Begelman1989})
	\begin{equation}
	\label{eq:Pressurebalance}
	h_{\rm j}\rho_{\rm j}c^2 (\Gamma \beta)_{\rm j,h}^2+P_{\rm j}=h_{\rm ej}\rho_{\rm ej}c^2 (\Gamma \beta)_{\rm h,ej}^2+P_{\rm ej}
	\end{equation}
	where $h,\rho, P$ are the dimensionless enthalpy, the density and the pressure of the fluid materials in their respective rest frames. The quantity $(\Gamma \beta)_{\rm j,h}$ ($(\Gamma \beta)_{\rm h,ej}$) measures the proper velocity of the jet (head) relative to the head (BNS merger ejecta).
	Assuming both the jet and the ejecta are initially cold, we can neglect the terms $P_{\rm j}, P_{\rm ej}$ in equation \ref{eq:Pressurebalance}. Using the Lorentz transformation we write $(\Gamma \beta)_{\rm x,y}=(\beta_{\rm x}-\beta_{\rm y})\Gamma_{\rm x}\Gamma_{\rm y}$. Plugging back into equation (\ref{eq:Pressurebalance}), we find
	\begin{equation}
	\label{eq:A2}
	h_{\rm j}\rho_{\rm j} \Gamma_{\rm j}^2   \beta_{\rm j}^2 \bigg(1-\frac{\beta_{\rm h}}{\beta_{\rm j}}\bigg)^2=h_{\rm ej}\rho_{\rm ej} \Gamma_{\rm ej}^2 (\beta_{\rm h}-\beta_{\rm ej})^2
	\end{equation}
	Noting that
	\begin{equation}
	\frac{4\pi r^2 h_{\rm j}\rho_{\rm j}\Gamma_{\rm j}^2 c^3}{4\pi r^2 h_{\rm ej}\rho_{\rm ej}\Gamma_{\rm ej}^2c^3}=\frac{L_{\rm e}\beta_{\rm ej}}{\dot{M}_{\rm ej}c^2 \Gamma_{\rm ej}h_{\rm ej}}=\frac{\tilde{L}}{\Gamma_{\rm ej}h_{\rm ej}}
	\end{equation}
	where in the last transition we have plugged in the definition of $\tilde{L}$ given in equation (\ref{eq:Ltilde}). Since for $\beta_{\rm ej}=0.25$, we have $(\Gamma_{\rm ej}h_{\rm ej})^{-1}\approx 1.05$ (where we have used an approximation for the enthalpy of monoenergetic relativistic gas, introduced by \citealt{Mathews1971}, according to which $h=1+\frac{1}{3}(1-\bar{\gamma}^{-2})$ where $\bar{\gamma}$ is the Lorentz factor of particles in the ejecta and is of the order of $\Gamma_{\rm ej}$), we can assume to a $\sim 5\%$ accuracy that $\tilde{L}\approx \frac{h_{\rm j}\rho_{\rm j}\Gamma_{\rm j}^2}{h_{\rm ej}\rho_{\rm ej}\Gamma_{\rm ej}^2}$.
	Plugging back into equation (\ref{eq:A2}), we find
	\begin{equation}
	\tilde{L}^{1/2}   \beta_{\rm j} \bigg(1-\frac{\beta_{\rm h}}{\beta_{\rm j}}\bigg)= (\beta_{\rm h}-\beta_{\rm ej})
	\end{equation}
	leading to 
	\begin{equation}
	\beta_{\rm h}=\frac{\beta_{\rm j}+\beta_{\rm ej}\tilde{L}^{-1/2}}{1+\tilde{L}^{-1/2}}
	\end{equation}
	which is the same as equation (\ref{eq:betah}).

	\section{A Monte Carlo estimation of the fraction of successful short GRB jets}
	\label{app:MC}
	As mentioned in \S \ref{sec:static}, GRBs with lower luminosities and shorter durations place the most stringent limits on the waiting time $t_{\rm w}$. 
	Here, we complement the qualitative discussion in \S \ref{sec:static} with a Monte Carlo (MC) estimation of the fraction of events  that result in successful sGRBs with a certain observed $\gamma$-ray luminosity $L_{\rm GRB}$, and duration $t_{\rm GRB}$. For the purposes of this calculation, we use our general jet breakout calculation that holds for a generic medium (i.e., not necessarily static or homologously expanding, see \S \ref{sec:general}). 
	
	As we are not interested in reproducing the exact  distribution of  \swift-BAT bursts in the $L_{\rm GRB}-t_{\rm GRB}$ parameter space, but rather aim to highlight the effect of the breakout, we employ the following method.
	We assume log-uniform priors\footnote{The limits on the distributions of $t_{\rm e}$, $L_{\rm e}$ do not affect the probability as long as (i) we reach sufficiently large $t_{\rm e}$, $L_{\rm e}$ such that the probability of breakout is essentially 100\% and  (ii) the adopted ranges are wide enough that all possible combinations of $t_{\rm e}$, $t_{\rm GRB}$ are realized.} for the engine time ($t_{\rm e}$) and engine power ($L_{\rm e}$) distributions and using equations (\ref{eq:tjb}), (\ref{eq:betah}), (\ref{eq:Ltilde}), (\ref{eq:homologous}), (\ref{eq:tjbhom}) we calculate the luminosity and duration of each jet that successfully breaks out from the BNS ejecta (i.e., with $t_{\rm e}>t_{\rm j,b}$).
	
\begin{figure}
		\centering
		\includegraphics[width=0.35\textwidth]{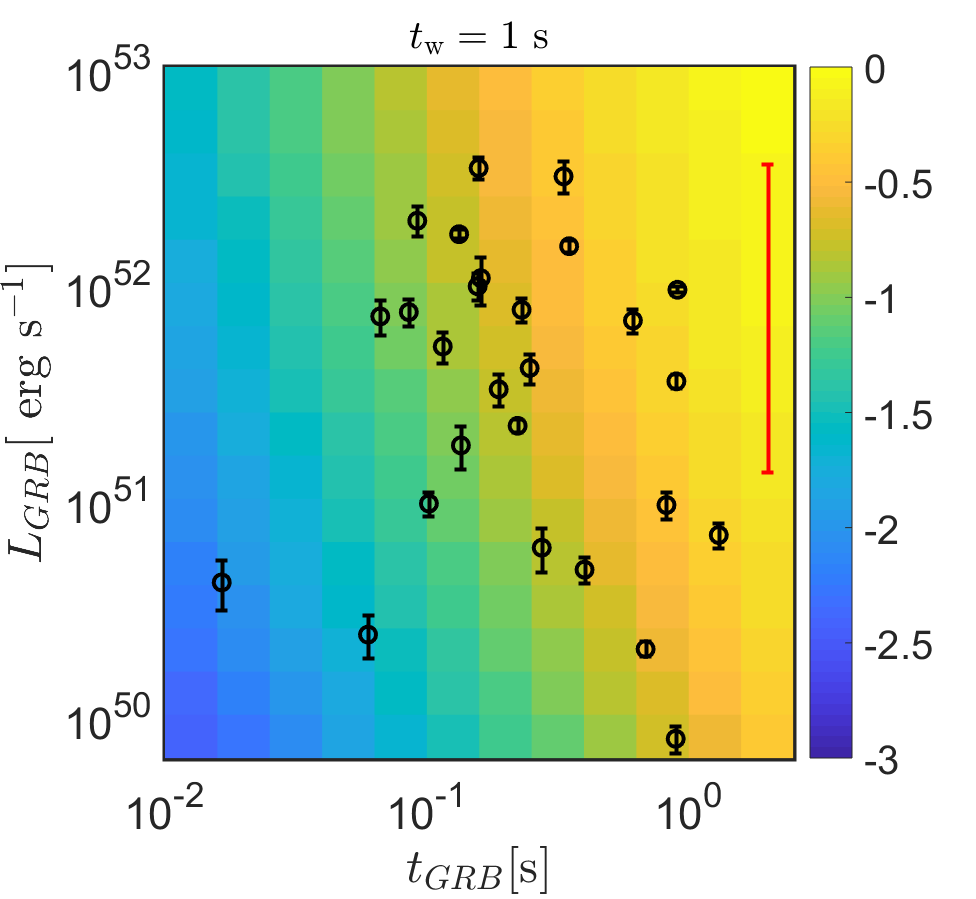}
		\includegraphics[width=0.35\textwidth]{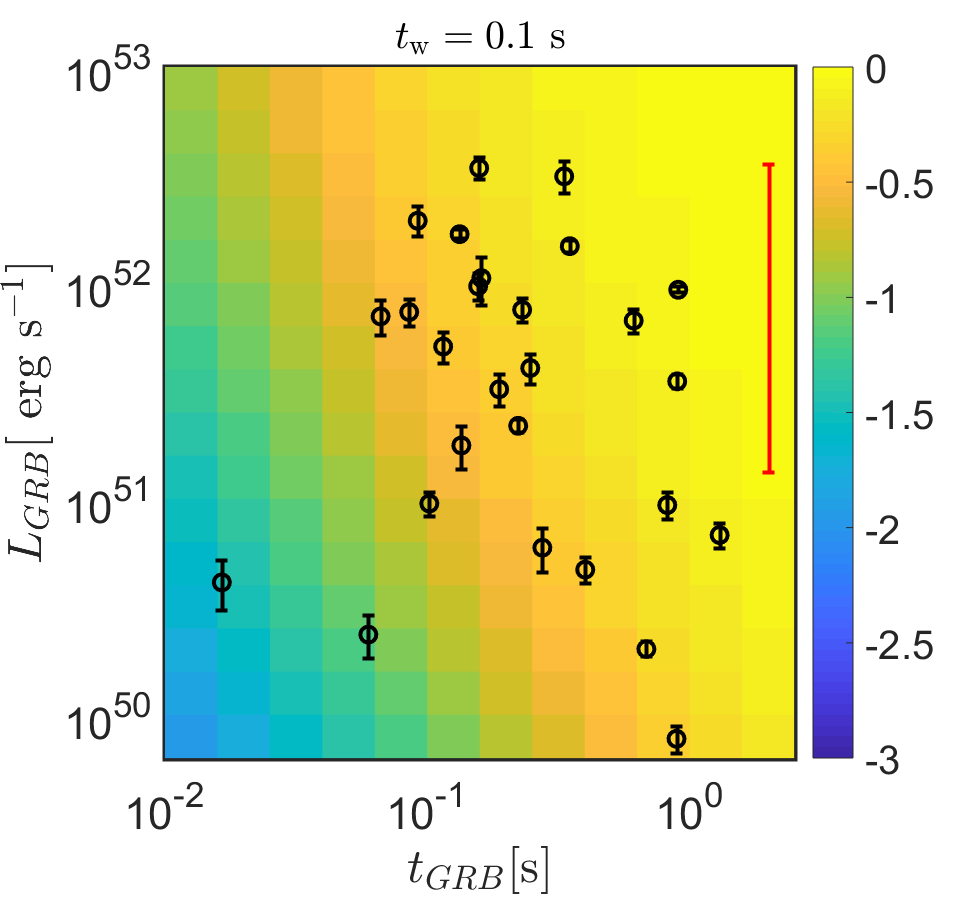}
		\caption{Density map of GRB luminosities and durations, as obtained from $10^8$ Monte Carlo realizations, assuming log-uniform priors for the engine time and engine power, and  different waiting times (marked on the plot). Colour denotes the probability of breakout (in logarithm). Black and red symbols have the same meaning as in Figure \ref{fig:twstatic}. For increasing values of $t_{\rm w}$, the probability of obtaining bursts with short duration and low GRB luminosity decreases.}\label{fig:numbers} 
	\end{figure}

	Figure \ref{fig:numbers} shows the density maps of all simulated bursts that successfully break out from the BNS ejecta, computed for two values of the waiting time ($t_{\rm w}=1$~s and 0.1~s) and using $10^8$ MC realizations for each case. \swift-BAT bursts with measured redshift (black circles) and GRB~170817 (red symbol) are overplotted.
	For a given value of $t_{\rm w}$, bursts with lower luminosity (or, equivalently engine power) have longer breakout times $t_{\rm j,b}$ (see e.g., Figure \ref{fig:tjbtw}). Thus, they are less likely to successfully break out from the BNS merger ejecta (they require longer engine activity durations, corresponding to a smaller fraction of simulated events). This results in a deficiency of simulated bursts with low $L_{\rm GRB}$. Furthermore, longer breakout times mean that a short GRB duration requires fine tuning in terms of $t_{\rm e}/t_{\rm j,b}$ (see equation \ref{eq:tGRB}). Therefore, the fraction of successful bursts decreases also with diminishing $t_{\rm GRB}$.
	
	For increasing values of $t_{\rm w}$, the probability of obtaining bursts with short duration and low GRB luminosity decreases (see top panel of Figure~\ref{fig:numbers}). This effect is largely insensitive to the assumed priors on the probabilities, as it is due to the low breakout probability. 
	As an example, for $t_{\rm w}=1$ s, there are 2/27 bursts that have breakout probabilities of $\sim 0.01$ and 9/27 with breakout probabilities $\lesssim 0.1$. These are well below the expected statistical fluctuations from Poisson statistics, corresponding to a $>5\sigma$ deviation. Of course, the exact probabilities depend on the assumed priors, but the overall conclusion, that low values of $t_{\rm w}$ are required to explain the observed sGRBs is largely insensitive of those priors.

	We finally note that the top right corners of both panels in Figure \ref{fig:numbers} (which for the adopted priors are expected to be the most populated) are in practice empty of \swift-BAT sGRBs.
	This result should not be surprising, as it simply reflects the fact that the true distributions of the engine properties (i.e., $t_{\rm e}$ and $L_{\rm e}$) are not expected to be as simple as the statistically independent log-uniform priors assumed here.
	
\begin{thebibliography}{}
		\expandafter\ifx\csname natexlab\endcsname\relax\def\natexlab#1{#1}\fi
		\providecommand{\url}[1]{\href{#1}{#1}}
		\providecommand{\dodoi}[1]{doi:~\href{http://doi.org/#1}{\nolinkurl{#1}}}
		\providecommand{\doeprint}[1]{\href{http://ascl.net/#1}{\nolinkurl{http://ascl.net/#1}}}
		\providecommand{\doarXiv}[1]{\href{https://arxiv.org/abs/#1}{\nolinkurl{https://arxiv.org/abs/#1}}}
		
		\bibitem[{Abbott {et~al.}(2017)}]{GW170817}
		Abbott, B.~P., {et~al.} 2017, Phys. Rev. Lett., 119, 161101,
		\dodoi{10.1103/PhysRevLett.119.161101}
		
		\bibitem[{{Aloy} {et~al.}(2005){Aloy}, {Janka}, \& {M{\"u}ller}}]{Aloy2005}
		{Aloy}, M.~A., {Janka}, H.~T., \& {M{\"u}ller}, E. 2005, \aap, 436, 273,
		\dodoi{10.1051/0004-6361:20041865}
		
		\bibitem[{{Band} {et~al.}(1993){Band}, {Matteson}, {Ford}, {Schaefer},
			{Palmer}, {Teegarden}, {Cline}, {Briggs}, {Paciesas}, {Pendleton}, {Fishman},
			{Kouveliotou}, {Meegan}, {Wilson}, \& {Lestrade}}]{Band1993}
		{Band}, D., {Matteson}, J., {Ford}, L., {et~al.} 1993, \apj, 413, 281,
		\dodoi{10.1086/172995}
		
		\bibitem[{{Bartos} {et~al.}(2019){Bartos}, {Lee}, {Corsi}, {M{\'a}rka}, \&
			{M{\'a}rka}}]{Bartos2019}
		{Bartos}, I., {Lee}, K.~H., {Corsi}, A., {M{\'a}rka}, Z., \& {M{\'a}rka}, S.
		2019, \mnras, 485, 4150, \dodoi{10.1093/mnras/stz719}
		
		\bibitem[{{Baumgarte} {et~al.}(2000){Baumgarte}, {Shapiro}, \&
			{Shibata}}]{Baumgarte2000ApJ...528L..29B}
		{Baumgarte}, T.~W., {Shapiro}, S.~L., \& {Shibata}, M. 2000, \apjl, 528, L29,
		\dodoi{10.1086/312425}
		
		\bibitem[{{Begelman} \& {Cioffi}(1989)}]{Begelman1989}
		{Begelman}, M.~C., \& {Cioffi}, D.~F. 1989, \apjl, 345, L21,
		\dodoi{10.1086/185542}
		
		\bibitem[{{Beniamini} {et~al.}(2018){Beniamini}, {Barniol Duran}, \&
			{Giannios}}]{BBDG2018}
		{Beniamini}, P., {Barniol Duran}, R., \& {Giannios}, D. 2018, \mnras, 476,
		1785, \dodoi{10.1093/mnras/sty340}
		
		\bibitem[{{Beniamini} {et~al.}(2020{\natexlab{a}}){Beniamini}, {Duque},
			{Daigne}, \& {Mochkovitch}}]{BDDM2019}
		{Beniamini}, P., {Duque}, R., {Daigne}, F., \& {Mochkovitch}, R.
		2020{\natexlab{a}}, \mnras, 492, 2847, \dodoi{10.1093/mnras/staa070}
		
		\bibitem[{{Beniamini} \& {Giannios}(2017)}]{BG2017}
		{Beniamini}, P., \& {Giannios}, D. 2017, \mnras, 468, 3202,
		\dodoi{10.1093/mnras/stx717}
		
		\bibitem[{{Beniamini} {et~al.}(2017){Beniamini}, {Giannios}, \&
			{Metzger}}]{BGM2017}
		{Beniamini}, P., {Giannios}, D., \& {Metzger}, B.~D. 2017, \mnras, 472, 3058,
		\dodoi{10.1093/mnras/stx2095}
		
		\bibitem[{{Beniamini} \& {Granot}(2016)}]{BG2016}
		{Beniamini}, P., \& {Granot}, J. 2016, \mnras, 459, 3635,
		\dodoi{10.1093/mnras/stw895}
		
		\bibitem[{{Beniamini} {et~al.}(2020{\natexlab{b}}){Beniamini}, {Granot}, \&
			{Gill}}]{Beniamini2020}
		{Beniamini}, P., {Granot}, J., \& {Gill}, R. 2020{\natexlab{b}}, \mnras,
		\dodoi{10.1093/mnras/staa538}
		
		\bibitem[{{Beniamini} \& {Nakar}(2019)}]{BN2019}
		{Beniamini}, P., \& {Nakar}, E. 2019, \mnras, 482, 5430,
		\dodoi{10.1093/mnras/sty3110}
		
		\bibitem[{{Beniamini} {et~al.}(2015){Beniamini}, {Nava}, {Duran}, \&
			{Piran}}]{Beniamini2015}
		{Beniamini}, P., {Nava}, L., {Duran}, R.~B., \& {Piran}, T. 2015, \mnras, 454,
		1073, \dodoi{10.1093/mnras/stv2033}
		
		\bibitem[{{Beniamini} {et~al.}(2019){Beniamini}, {Petropoulou}, {Barniol
				Duran}, \& {Giannios}}]{Beniamini2019}
		{Beniamini}, P., {Petropoulou}, M., {Barniol Duran}, R., \& {Giannios}, D.
		2019, \mnras, 483, 840, \dodoi{10.1093/mnras/sty3093}
		
		\bibitem[{{Beniamini} \& {Piran}(2016)}]{BP2016}
		{Beniamini}, P., \& {Piran}, T. 2016, \mnras, 456, 4089,
		\dodoi{10.1093/mnras/stv2903}
		
		\bibitem[{{Berger}(2014)}]{Berger2014}
		{Berger}, E. 2014, \araa, 52, 43, \dodoi{10.1146/annurev-astro-081913-035926}
		
		\bibitem[{{Bloom} {et~al.}(2006){Bloom}, {Prochaska}, {Pooley}, {Blake},
			{Foley}, {Jha}, {Ramirez-Ruiz}, {Granot}, {Filippenko}, {Sigurdsson},
			{Barth}, {Chen}, {Cooper}, {Falco}, {Gal}, {Gerke}, {Gladders}, {Greene},
			{Hennanwi}, {Ho}, {Hurley}, {Koester}, {Li}, {Lubin}, {Newman}, {Perley},
			{Squires}, \& {Wood-Vasey}}]{Bloom2006}
		{Bloom}, J.~S., {Prochaska}, J.~X., {Pooley}, D., {et~al.} 2006, \apj, 638,
		354, \dodoi{10.1086/498107}
		
		\bibitem[{{Bromberg} {et~al.}(2011){Bromberg}, {Nakar}, {Piran}, \&
			{Sari}}]{Bromberg2011}
		{Bromberg}, O., {Nakar}, E., {Piran}, T., \& {Sari}, R. 2011, \apj, 740, 100,
		\dodoi{10.1088/0004-637X/740/2/100}
		
		\bibitem[{{Bromberg} {et~al.}(2012){Bromberg}, {Nakar}, {Piran}, \&
			{Sari}}]{Bromberg2012}
		---. 2012, \apj, 749, 110, \dodoi{10.1088/0004-637X/749/2/110}
		
		\bibitem[{{Bromberg} {et~al.}(2013){Bromberg}, {Nakar}, {Piran}, \&
			{Sari}}]{Bromberg2013}
		---. 2013, \apj, 764, 179, \dodoi{10.1088/0004-637X/764/2/179}
		
		\bibitem[{{Burns} {et~al.}(2018){Burns}, {Veres}, {Connaughton}, {Racusin},
			{Briggs}, {Christensen}, {Goldstein}, {Hamburg}, {Kocevski}, {McEnery},
			{Bissaldi}, {Dal Canton}, {Cleveland}, {Gibby}, {Hui}, {von Kienlin},
			{Mailyan}, {Paciesas}, {Roberts}, {Siellez}, {Stanbro}, \&
			{Wilson-Hodge}}]{Burns2018}
		{Burns}, E., {Veres}, P., {Connaughton}, V., {et~al.} 2018, \apjl, 863, L34,
		\dodoi{10.3847/2041-8213/aad813}
		
		\bibitem[{{Christie} {et~al.}(2019){Christie}, {Lalakos}, {Tchekhovskoy},
			{Fern{\'a}ndez}, {Foucart}, {Quataert}, \& {Kasen}}]{Christie2019}
		{Christie}, I.~M., {Lalakos}, A., {Tchekhovskoy}, A., {et~al.} 2019, \mnras,
		\dodoi{10.1093/mnras/stz2552}
		
		\bibitem[{{Daigne} \& {Mochkovitch}(1998)}]{DM1998}
		{Daigne}, F., \& {Mochkovitch}, R. 1998, \mnras, 296, 275,
		\dodoi{10.1046/j.1365-8711.1998.01305.x}
		
		\bibitem[{{Dichiara} {et~al.}(2020){Dichiara}, {Troja}, {O'Connor}, {Marshall},
			{Beniamini}, {Cannizzo}, {Lien}, \& {Sakamoto}}]{Dichiara2020}
		{Dichiara}, S., {Troja}, E., {O'Connor}, B., {et~al.} 2020, \mnras, 492, 5011,
		\dodoi{10.1093/mnras/staa124}
		
		\bibitem[{{Duffell} {et~al.}(2018){Duffell}, {Quataert}, {Kasen}, \&
			{Klion}}]{Duffell2018}
		{Duffell}, P.~C., {Quataert}, E., {Kasen}, D., \& {Klion}, H. 2018, \apj, 866,
		3, \dodoi{10.3847/1538-4357/aae084}
		
		\bibitem[{{Favata}(2014)}]{Favata2014}
		{Favata}, M. 2014, \prl, 112, 101101, \dodoi{10.1103/PhysRevLett.112.101101}
		
		\bibitem[{{Finstad} {et~al.}(2018){Finstad}, {De}, {Brown}, {Berger}, \&
			{Biwer}}]{Finstad2018}
		{Finstad}, D., {De}, S., {Brown}, D.~A., {Berger}, E., \& {Biwer}, C.~M. 2018,
		\apjl, 860, L2, \dodoi{10.3847/2041-8213/aac6c1}
		
		\bibitem[{{Flanagan} \& {Hinderer}(2008)}]{FH2008}
		{Flanagan}, {\'E}.~{\'E}., \& {Hinderer}, T. 2008, \prd, 77, 021502,
		\dodoi{10.1103/PhysRevD.77.021502}
		
		\bibitem[{{Fryer} {et~al.}(2015){Fryer}, {Belczynski}, {Ramirez-Ruiz},
			{Rosswog}, {Shen}, \& {Steiner}}]{Fryer2015}
		{Fryer}, C.~L., {Belczynski}, K., {Ramirez-Ruiz}, E., {et~al.} 2015, \apj, 812,
		24, \dodoi{10.1088/0004-637X/812/1/24}
		
		\bibitem[{{Gehrels} {et~al.}(2004){Gehrels}, {Chincarini}, {Giommi}, {Mason},
			{Nousek}, {Wells}, {White}, {Barthelmy}, {Burrows}, {Cominsky}, {Hurley},
			{Marshall}, {M{\'e}sz{\'a}ros}, {Roming}, {Angelini}, {Barbier}, {Belloni},
			{Campana}, {Caraveo}, {Chester}, {Citterio}, {Cline}, {Cropper}, {Cummings},
			{Dean}, {Feigelson}, {Fenimore}, {Frail}, {Fruchter}, {Garmire}, {Gendreau},
			{Ghisellini}, {Greiner}, {Hill}, {Hunsberger}, {Krimm}, {Kulkarni}, {Kumar},
			{Lebrun}, {Lloyd-Ronning}, {Markwardt}, {Mattson}, {Mushotzky}, {Norris},
			{Osborne}, {Paczynski}, {Palmer}, {Park}, {Parsons}, {Paul}, {Rees},
			{Reynolds}, {Rhoads}, {Sasseen}, {Schaefer}, {Short}, {Smale}, {Smith},
			{Stella}, {Tagliaferri}, {Takahashi}, {Tashiro}, {Townsley}, {Tueller},
			{Turner}, {Vietri}, {Voges}, {Ward}, {Willingale}, {Zerbi}, \&
			{Zhang}}]{2004ApJ...611.1005G}
		{Gehrels}, N., {Chincarini}, G., {Giommi}, P., {et~al.} 2004, \apj, 611, 1005,
		\dodoi{10.1086/422091}
		
		\bibitem[{{Geng} {et~al.}(2019){Geng}, {Zhang}, {K{\"o}lligan}, {Kuiper}, \&
			{Huang}}]{Geng2019}
		{Geng}, J.-J., {Zhang}, B., {K{\"o}lligan}, A., {Kuiper}, R., \& {Huang}, Y.-F.
		2019, \apjl, 877, L40, \dodoi{10.3847/2041-8213/ab224b}
		
		\bibitem[{{Giannios}(2012)}]{Giannios2012}
		{Giannios}, D. 2012, \mnras, 422, 3092,
		\dodoi{10.1111/j.1365-2966.2012.20825.x}
		
		\bibitem[{{Gill} {et~al.}(2019{\natexlab{a}}){Gill}, {Granot}, {De Colle}, \&
			{Urrutia}}]{Gill+19}
		{Gill}, R., {Granot}, J., {De Colle}, F., \& {Urrutia}, G. 2019{\natexlab{a}},
		\apj, 883, 15, \dodoi{10.3847/1538-4357/ab3577}
		
		\bibitem[{{Gill} {et~al.}(2019{\natexlab{b}}){Gill}, {Nathanail}, \&
			{Rezzolla}}]{GNR2019}
		{Gill}, R., {Nathanail}, A., \& {Rezzolla}, L. 2019{\natexlab{b}}, \apj, 876,
		139, \dodoi{10.3847/1538-4357/ab16da}
		
		\bibitem[{{Goldstein} {et~al.}(2017){Goldstein}, {Veres}, {Burns}, {Briggs},
			{Hamburg}, {Kocevski}, {Wilson-Hodge}, {Preece}, {Poolakkil}, {Roberts},
			{Hui}, {Connaughton}, {Racusin}, {von Kienlin}, {Dal Canton}, {Christensen},
			{Littenberg}, {Siellez}, {Blackburn}, {Broida}, {Bissaldi}, {Cleveland},
			{Gibby}, {Giles}, {Kippen}, {McBreen}, {McEnery}, {Meegan}, {Paciesas}, \&
			{Stanbro}}]{Goldstein2017}
		{Goldstein}, A., {Veres}, P., {Burns}, E., {et~al.} 2017, \apjl, 848, L14,
		\dodoi{10.3847/2041-8213/aa8f41}
		
		\bibitem[{{Granot} {et~al.}(2017){Granot}, {Guetta}, \&
			{Gill}}]{Granot2017ApJ...850L..24G}
		{Granot}, J., {Guetta}, D., \& {Gill}, R. 2017, \apjl, 850, L24,
		\dodoi{10.3847/2041-8213/aa991d}
		
		\bibitem[{{Hamidani} {et~al.}(2020){Hamidani}, {Kiuchi}, \&
			{Ioka}}]{Hamidani2020}
		{Hamidani}, H., {Kiuchi}, K., \& {Ioka}, K. 2020, \mnras, 491, 3192,
		\dodoi{10.1093/mnras/stz3231}
		
		\bibitem[{{Harrison} {et~al.}(2018){Harrison}, {Gottlieb}, \&
			{Nakar}}]{Harrison2018}
		{Harrison}, R., {Gottlieb}, O., \& {Nakar}, E. 2018, \mnras, 477, 2128,
		\dodoi{10.1093/mnras/sty760}
		
		\bibitem[{{Hotokezaka} {et~al.}(2018){Hotokezaka}, {Beniamini}, \&
			{Piran}}]{Hotokezaka2018}
		{Hotokezaka}, K., {Beniamini}, P., \& {Piran}, T. 2018, International Journal
		of Modern Physics D, 27, 1842005, \dodoi{10.1142/S0218271818420051}
		
		\bibitem[{{Just} {et~al.}(2016){Just}, {Obergaulinger}, {Janka}, {Bauswein}, \&
			{Schwarz}}]{Just2016}
		{Just}, O., {Obergaulinger}, M., {Janka}, H.~T., {Bauswein}, A., \& {Schwarz},
		N. 2016, \apjl, 816, L30, \dodoi{10.3847/2041-8205/816/2/L30}
		
		\bibitem[{{Kasen} {et~al.}(2017){Kasen}, {Metzger}, {Barnes}, {Quataert}, \&
			{Ramirez-Ruiz}}]{Kasen2017}
		{Kasen}, D., {Metzger}, B., {Barnes}, J., {Quataert}, E., \& {Ramirez-Ruiz}, E.
		2017, \nat, 551, 80, \dodoi{10.1038/nature24453}
		
		\bibitem[{{Kathirgamaraju} {et~al.}(2018){Kathirgamaraju}, {Barniol Duran}, \&
			{Giannios}}]{Kathirgamaraju2018}
		{Kathirgamaraju}, A., {Barniol Duran}, R., \& {Giannios}, D. 2018, \mnras, 473,
		L121, \dodoi{10.1093/mnrasl/slx175}
		
		\bibitem[{{Kathirgamaraju} {et~al.}(2019){Kathirgamaraju}, {Tchekhovskoy},
			{Giannios}, \& {Barniol Duran}}]{Kathirgamaraju2019}
		{Kathirgamaraju}, A., {Tchekhovskoy}, A., {Giannios}, D., \& {Barniol Duran},
		R. 2019, \mnras, 484, L98, \dodoi{10.1093/mnrasl/slz012}
		
		\bibitem[{{Kumar} \& {Narayan}(2009)}]{KN2009}
		{Kumar}, P., \& {Narayan}, R. 2009, \mnras, 395, 472,
		\dodoi{10.1111/j.1365-2966.2009.14539.x}
		
		\bibitem[{{Lamb} \& {Kobayashi}(2017)}]{lamb2017}
		{Lamb}, G.~P., \& {Kobayashi}, S. 2017, \mnras, 472, 4953,
		\dodoi{10.1093/mnras/stx2345}
		
		\bibitem[{{Lazzati} {et~al.}(2017){Lazzati}, {Deich}, {Morsony}, \&
			{Workman}}]{Lazzati2017}
		{Lazzati}, D., {Deich}, A., {Morsony}, B.~J., \& {Workman}, J.~C. 2017, \mnras,
		471, 1652, \dodoi{10.1093/mnras/stx1683}
		
		\bibitem[{{Lazzati} \& {Perna}(2019)}]{Lazzati2019}
		{Lazzati}, D., \& {Perna}, R. 2019, \apj, 881, 89,
		\dodoi{10.3847/1538-4357/ab2e06}
		
		\bibitem[{{Lyutikov}(2020)}]{Lyutikov2020}
		{Lyutikov}, M. 2020, \mnras, 491, 483, \dodoi{10.1093/mnras/stz3044}
		
		\bibitem[{{Mandhai} {et~al.}(2019){Mandhai}, {Tanvir}, {Lamb}, {Levan}, \&
			{Tsang}}]{Mandhai2019}
		{Mandhai}, S., {Tanvir}, N., {Lamb}, G., {Levan}, A., \& {Tsang}, D. 2019,
		arXiv e-prints, arXiv:1908.00100.
		\newblock \doarXiv{1908.00100}
		
		\bibitem[{{Marti} {et~al.}(1994){Marti}, {Mueller}, \& {Ibanez}}]{Marti1994}
		{Marti}, J.~M., {Mueller}, E., \& {Ibanez}, J.~M. 1994, \aap, 281, L9
		
		\bibitem[{{Mathews}(1971)}]{Mathews1971}
		{Mathews}, W.~G. 1971, \apj, 165, 147, \dodoi{10.1086/150883}
		
		\bibitem[{{Matzner}(2003)}]{Matzner2003}
		{Matzner}, C.~D. 2003, \mnras, 345, 575,
		\dodoi{10.1046/j.1365-8711.2003.06969.x}
		
		\bibitem[{{Metzger} {et~al.}(2018){Metzger}, {Beniamini}, \&
			{Giannios}}]{MBG2018}
		{Metzger}, B.~D., {Beniamini}, P., \& {Giannios}, D. 2018, \apj, 857, 95,
		\dodoi{10.3847/1538-4357/aab70c}
		
		\bibitem[{{Metzger} {et~al.}(2011){Metzger}, {Giannios}, {Thompson},
			{Bucciantini}, \& {Quataert}}]{Metzger2011}
		{Metzger}, B.~D., {Giannios}, D., {Thompson}, T.~A., {Bucciantini}, N., \&
		{Quataert}, E. 2011, \mnras, 413, 2031,
		\dodoi{10.1111/j.1365-2966.2011.18280.x}
		
		\bibitem[{{Mizuta} \& {Ioka}(2013)}]{Mizuta2013}
		{Mizuta}, A., \& {Ioka}, K. 2013, \apj, 777, 162,
		\dodoi{10.1088/0004-637X/777/2/162}
		
		\bibitem[{{Moharana} \& {Piran}(2017)}]{Moharana2017}
		{Moharana}, R., \& {Piran}, T. 2017, \mnras, 472, L55,
		\dodoi{10.1093/mnrasl/slx131}
		
		\bibitem[{{Mooley} {et~al.}(2018){Mooley}, {Deller}, {Gottlieb}, {Nakar},
			{Hallinan}, {Bourke}, {Frail}, {Horesh}, {Corsi}, \&
			{Hotokezaka}}]{Mooley2018}
		{Mooley}, K.~P., {Deller}, A.~T., {Gottlieb}, O., {et~al.} 2018, \nat, 561,
		355, \dodoi{10.1038/s41586-018-0486-3}
		
		\bibitem[{{Murguia-Berthier} {et~al.}(2014){Murguia-Berthier}, {Montes},
			{Ramirez-Ruiz}, {De Colle}, \& {Lee}}]{MB2014}
		{Murguia-Berthier}, A., {Montes}, G., {Ramirez-Ruiz}, E., {De Colle}, F., \&
		{Lee}, W.~H. 2014, \apjl, 788, L8, \dodoi{10.1088/2041-8205/788/1/L8}
		
		\bibitem[{{Murguia-Berthier} {et~al.}(2017){Murguia-Berthier}, {Ramirez-Ruiz},
			{Montes}, {De Colle}, {Rezzolla}, {Rosswog}, {Takami}, {Perego}, \&
			{Lee}}]{MB2017}
		{Murguia-Berthier}, A., {Ramirez-Ruiz}, E., {Montes}, G., {et~al.} 2017, \apjl,
		835, L34, \dodoi{10.3847/2041-8213/aa5b9e}
		
		\bibitem[{{Nagakura} {et~al.}(2014){Nagakura}, {Hotokezaka}, {Sekiguchi},
			{Shibata}, \& {Ioka}}]{Nagakura2014}
		{Nagakura}, H., {Hotokezaka}, K., {Sekiguchi}, Y., {Shibata}, M., \& {Ioka}, K.
		2014, \apjl, 784, L28, \dodoi{10.1088/2041-8205/784/2/L28}
		
		\bibitem[{{Nakar}(2007)}]{Nakar2007}
		{Nakar}, E. 2007, \physrep, 442, 166, \dodoi{10.1016/j.physrep.2007.02.005}
		
		\bibitem[{{Nakar}(2019)}]{Nakar2019}
		---. 2019, arXiv e-prints, arXiv:1912.05659.
		\newblock \doarXiv{1912.05659}
		
		\bibitem[{{Nakar} \& {Sari}(2010)}]{NakarSari2010}
		{Nakar}, E., \& {Sari}, R. 2010, \apj, 725, 904,
		\dodoi{10.1088/0004-637X/725/1/904}
		
		\bibitem[{{Nava} {et~al.}(2011){Nava}, {Ghirlanda}, {Ghisellini}, \&
			{Celotti}}]{Nava2011}
		{Nava}, L., {Ghirlanda}, G., {Ghisellini}, G., \& {Celotti}, A. 2011, \aap,
		530, A21, \dodoi{10.1051/0004-6361/201016270}
		
		\bibitem[{{Oganesyan} {et~al.}(2019){Oganesyan}, {Ascenzi}, {Branchesi},
			{Sharan Salafia}, {Dall'Osso}, \& {Ghirlanda}}]{Oganesyan2019}
		{Oganesyan}, G., {Ascenzi}, S., {Branchesi}, M., {et~al.} 2019, arXiv e-prints,
		arXiv:1904.08786.
		\newblock \doarXiv{1904.08786}
		
		\bibitem[{{Petropoulou} {et~al.}(2017){Petropoulou}, {Barniol Duran}, \&
			{Giannios}}]{Petropoulou2017}
		{Petropoulou}, M., {Barniol Duran}, R., \& {Giannios}, D. 2017, \mnras, 472,
		2722, \dodoi{10.1093/mnras/stx2151}
		
		\bibitem[{{Radice} {et~al.}(2018){Radice}, {Perego}, {Zappa}, \&
			{Bernuzzi}}]{Radice2018}
		{Radice}, D., {Perego}, A., {Zappa}, F., \& {Bernuzzi}, S. 2018, \apjl, 852,
		L29, \dodoi{10.3847/2041-8213/aaa402}
		
		\bibitem[{{Ramirez-Ruiz} {et~al.}(2002){Ramirez-Ruiz}, {Celotti}, \&
			{Rees}}]{RamirezRuiz2002}
		{Ramirez-Ruiz}, E., {Celotti}, A., \& {Rees}, M.~J. 2002, \mnras, 337, 1349,
		\dodoi{10.1046/j.1365-8711.2002.05995.x}
		
		\bibitem[{{Salafia} {et~al.}(2019){Salafia}, {Barbieri}, {Ascenzi}, \&
			{Toffano}}]{Salafia2019}
		{Salafia}, O.~S., {Barbieri}, C., {Ascenzi}, S., \& {Toffano}, M. 2019, arXiv
		e-prints, arXiv:1907.07599.
		\newblock \doarXiv{1907.07599}
		
		\bibitem[{{Sari} \& {Piran}(1997)}]{SP1997}
		{Sari}, R., \& {Piran}, T. 1997, \apj, 485, 270, \dodoi{10.1086/304428}
		
		\bibitem[{{Sobacchi} {et~al.}(2017){Sobacchi}, {Granot}, {Bromberg}, \&
			{Sormani}}]{Sobacchi2017}
		{Sobacchi}, E., {Granot}, J., {Bromberg}, O., \& {Sormani}, M.~C. 2017, \mnras,
		472, 616, \dodoi{10.1093/mnras/stx2083}
		
		\bibitem[{{Troja} {et~al.}(2018){Troja}, {Ryan}, {Piro}, {van Eerten}, {Cenko},
			{Yoon}, {Lee}, {Im}, {Sakamoto}, {Gatkine}, {Kutyrev}, \&
			{Veilleux}}]{Troja2018}
		{Troja}, E., {Ryan}, G., {Piro}, L., {et~al.} 2018, Nature Communications, 9,
		4089, \dodoi{10.1038/s41467-018-06558-7}
		
		\bibitem[{{Troja} {et~al.}(2019){Troja}, {van Eerten}, {Ryan}, {Ricci},
			{Burgess}, {Wieringa}, {Piro}, {Cenko}, \& {Sakamoto}}]{Troja2019}
		{Troja}, E., {van Eerten}, H., {Ryan}, G., {et~al.} 2019, \mnras, 489, 1919,
		\dodoi{10.1093/mnras/stz2248}
		
		\bibitem[{{Wanderman} \& {Piran}(2015)}]{WP2015}
		{Wanderman}, D., \& {Piran}, T. 2015, \mnras, 448, 3026,
		\dodoi{10.1093/mnras/stv123}
		
		\bibitem[{{Xie} {et~al.}(2018){Xie}, {Zrake}, \& {MacFadyen}}]{Xie2018}
		{Xie}, X., {Zrake}, J., \& {MacFadyen}, A. 2018, \apj, 863, 58,
		\dodoi{10.3847/1538-4357/aacf9c}
		
		\bibitem[{{Zhang}(2019)}]{Zhang2019}
		{Zhang}, B. 2019, Frontiers of Physics, 14, 64402,
		\dodoi{10.1007/s11467-019-0913-4}
		
		\bibitem[{{Zhang} \& {Yan}(2011)}]{Zhang2011}
		{Zhang}, B., \& {Yan}, H. 2011, \apj, 726, 90,
		\dodoi{10.1088/0004-637X/726/2/90}
		
	\end{thebibliography}
\end{document}